\documentclass[preprint,authoryear]{elsarticle}

\usepackage[latin1]{inputenc}
\usepackage[english]{babel} 
\usepackage{subfigure}
\usepackage{amsmath,amsfonts}
\usepackage{setspace}
\usepackage{slashbox} 
\usepackage{tabularx} 
\usepackage[normalem]{ulem} 

\allowdisplaybreaks[4] 
\newcommand{\R}{\textit{I\!R}}
\newcommand{\tablesize}{\fontsize{8}{8}\selectfont}

\usepackage{amssymb}
\usepackage{amsthm}

\begin{document}

\begin{frontmatter}



\title{Bartlett corrections in beta regression models}

 \author[label1]{F.~M.~Bayer\corref{cor1}}
 \ead{bayer@ufsm.br}
 \address[label1]{Departamento de Estatística, Universidade Federal de Santa Maria}
 \cortext[cor1]{Principal corresponding author}  
 
 \author[label2]{F.~Cribari-Neto\corref{cor2}}
 \ead{cribari@de.ufpe.br}
 \address[label2]{Departamento de Estatística, Universidade Federal de Pernambuco}
 \cortext[cor2]{Corresponding author}
 


\begin{abstract}
We consider the issue of performing accurate small-sample testing inference in beta regression models, which are useful for modeling continuous variates that assume values in $(0,1)$, such as rates and proportions. We derive the Bartlett correction to the likelihood ratio test statistic and also consider a bootstrap Bartlett correction. Using Monte Carlo simulations we compare the finite sample performances of the two corrected tests to that of the standard likelihood ratio test and also to its variant that employs Skovgaard's adjustment; the latter is already available in the literature. The numerical evidence favors the corrected tests we propose. We also present an empirical application.

\end{abstract}


\end{frontmatter}


\section{Introduction}

Regression analysis is commonly used to model the relationship between a dependent variable (response) and a set of explanatory variables (covariates). The linear regression model is the most used regression model in empirical applications, but it is not appropriate when the variable of interest assume values in the standard unit interval, as is the case of rates and proportions. For these situations \cite{Ferrari2004} proposed a regression model based on the assumption that the response ($y$) is beta distributed. Their model is similar to those that belong to the class of generalized linear models \citep{McCullagh1989}. 

\begin{figure}[htp]
\subfigure[$\;\phi=10$]{\label{F:densities_10}  \includegraphics[width=0.495\textwidth]{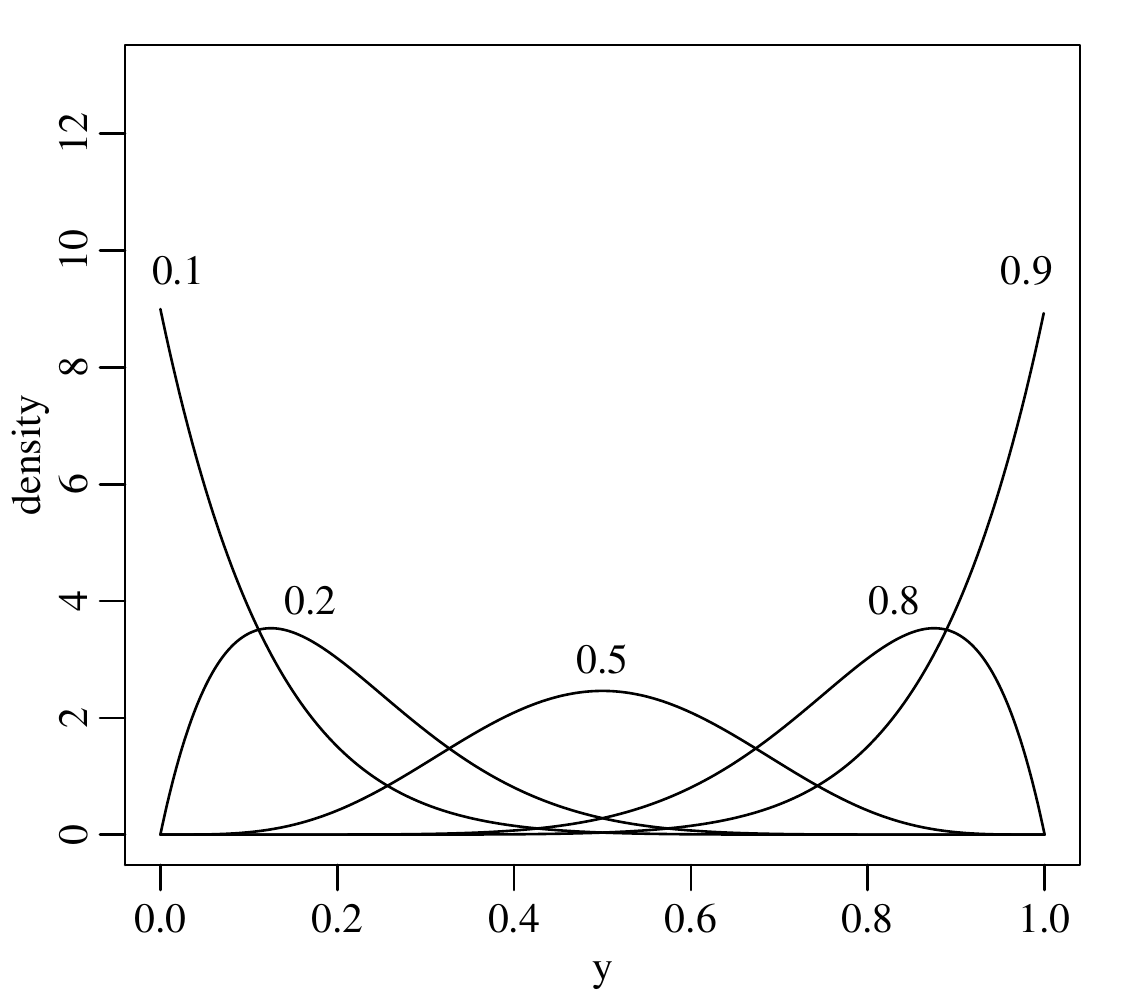}}
\subfigure[$\;\phi=90$]{\label{F:densities_90}  \includegraphics[width=0.495\textwidth] {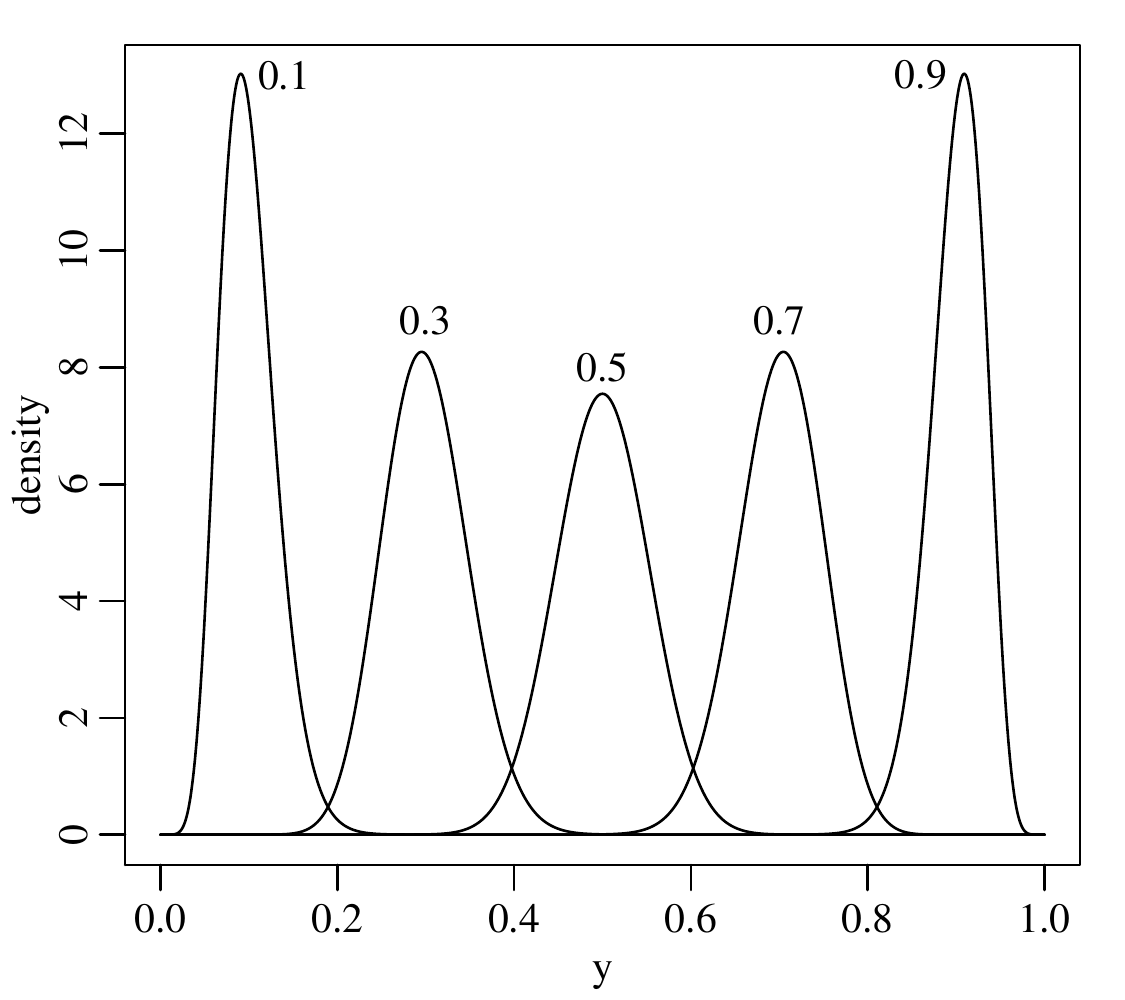}}
\caption{Beta densities for different values of $\mu$ (indicated in the panels), with $\phi=10$ (a) and $\phi=90$ (b).}\label{F:densities_beta}
\end{figure}

The beta density can be expressed as 
\begin{equation}\label{E:density}
f(y;\mu,\phi)=\frac{\Gamma{(\phi)}}{\Gamma{(\mu \phi)}\Gamma{((1-\mu)\phi)}}y^{\,\mu \,\phi -1}(1-y)^{(1-\mu)\phi-1}, \; \; 0<y<1,
\end{equation}
and
\begin{equation*}
{\rm E}(y)=\mu, \quad {\rm var}(y)=\frac{V(\mu)}{1+\phi},
\end{equation*}
where $V(\mu)\!=\mu\,(1\!-\mu)$ is the variance function and $\phi$ can be viewed as a precision parameter. The beta distribution is very flexible since its density can assume different shapes depending on the values of the two parameters. In particular, it can be symmetric, asymmetric, J-shaped and inverted J-shaped; see Figure~\ref{F:densities_beta}.

The class of beta regression models allows practitioners to model responses that belong to the interval $(0,1)$ using a regression structure that contains a link function, covariates and unknown parameters. Several authors have used beta regression models and alternative modeling strategies in different fields; see, e.g., \cite{Brehm1993}, \cite{Hancox2010}, \cite{Kieschnick2003}, \cite{Smithson2006} and \cite{Zucco2008}. 

One may be tempted to view the logistic regression as an alternative to the class of beta regressions. However, 
logistic regression is used when the response is binary, i.e., $y$ only assumes two values, namely: 0 and 1. In that case, one models $\Pr(y = 1)$ as a function of covariates. Beta regression, on the other hand, is used when the response is continuous and assume values in the standard unit interval. Beta regression is useful for modeling rates, proportions, concentration indices (e.g., Gini) and other variates that assume values in $(0,1)$ or, more generally, in $(a, b)$, where $a$ and $b$ are known ($a < b$). 

Testing inference in beta regression is usually carried out using the likelihood ratio test. The test employs an approximate critical value which is obtained from the test statistic limiting null distribution ($\chi^2$). It is thus an approximate test and size distortions are likely to take place in small samples. This happens because when the number of data points is not large the test statistic exact null distribution is oftentimes poorly approximated by its asymptotic counterpart. Testing inference can be made more reliable by transforming the likelihood ratio statistic using a Bartlett correction \citep{Bartlett1937}. Such a correction depends on the log-likelihood cumulants and mixed cumulants up to fourth order. The derivation of a closed-form expression for the Bartlett correction factor in beta regressions can be quite cumbersome since the mean and precision parameters are not orthogonal, unlike generalized linear models. 

A useful approach to improve inferences in small samples, particularly when the Bartlett correction is analytically cumbersome, is Skovgaard's adjustment \citep{Skovgaard2001}. This adjustment is more straightforward than the Bartlett correction, only requiring second order log-likelihood derivatives. It does not require orthogonality between nuisance parameters and parameters of interest. Skovgaard's adjustment for varying dispersion beta and inflated beta regressions were derived by \cite{Pinheiro2011} and \cite{Pereira2010}, respectively. \cite{Ferrari2008} obtained a similar adjustment for exponential family nonlinear models. The numerical results presented by these autors reveal that the modified likelihood ratio test obtained using Skovgaard's proposal is less size distorted than the original likelihood ratio test when the sample size is small.  

A shortcoming of Skovgaard's correction is that it does not improve the rate at which size distortions vanish, i.e., it does not yield asymptotic refinements. As noted earlier, however, Bartlett corrections are more difficult to obtain. They deliver asymptotic refinements and are usually derived using a general result given by \cite{Lawley1956}. An alternative is to use results in \cite{Cordeiro1993} which are written matrix fashion. Another alternative for models in which the derivation of Bartlett correction is analytically cumbersome is the bootstrap Bartlett correction \citep{Rocke1989}. Here, the Bartlett correction factor is determined using bootstrap resampling \citep{Efron1979}.

Our main goal in this paper is to derive the Bartlett correction factor to the likelihood ratio test in the class of beta regressions. The derivation is quite cumbersome since in beta regressions the mean regression parameter vector is not orthogonal to the precision parameter. We were able to obtain, after extensive algebra, the Bartlett correction for fixed dispersion beta regressions. We also consider the bootstrap Bartlett correction, i.e., we numerically estimate the Bartlett correction factor. Finally, we perform extensive Monte Carlo simulations where we compare the finite sample behavior of Bartlett corrected tests (analytically and numerically) to that of the modified likelihood ratio test of \cite{Pinheiro2011}. The numerical evidence favors the two Bartlett corrected tests, especially the bootstrap Bartlett corrected test. 

The paper unfolds as follows. Section~\ref{S:model} introduces the beta regression model proposed by \cite{Ferrari2004}. In Section~\ref{S:correction} we derive the Bartlett correction factor to the likelihood ratio test in fixed dispersion beta regressions. We also present the bootstrap Bartlett correction and the modified likelihood ratio statistics obtained by \cite{Pinheiro2011}. Monte Carlo Simulation results are presented and discussed in Section~\ref{S:simulation}. Section~\ref{S:application} presents an application that uses real (not simulated) data. Concluding remarks are offered in the last section and the log-likelihood cumulants we derived are presented in the \ref{A:cumulants}.

\section{The beta regression model}\label{S:model}

Let $y=(y_1,\ldots,y_n)^{\top}$ be a vector of $n$ independent random variables, each $y_i$, $i=1,\ldots,n$, having density \eqref{E:density} with mean $\mu_i$ and unknown parameter precision $\phi$. The beta regression model can be written as
\begin{equation}\label{E:g_mu}
g(\mu_i)=\sum_{j=1}^{p}x_{ij}\beta_j=\eta_i,
\end{equation}
where ${\beta}=(\beta_1,\ldots,\beta_p)^{\top}$ is an unknown vector parameter and $x_{i1},\ldots,x_{ip}$ are observations on the $p$ covariates ($p<n$). When an intercept is included in the model, we have $x_{i1}\!=\!1$, for $i\!=\!1,\ldots,n$. Finally, $g(\cdot)$ is a strictly monotonic and twice differentiable link function, with domain in $(0, 1)$ and image in \R. Some commonly used link functions are logit, probit, cloglog, loglog and Cauchy.

Estimation of the $k$-dimensional parameter vector $\theta=(\beta^{\top},\phi)^{\top}$, where $k=(p+1)$, can be performed by maximum likelihood. The log-likelihood function is
\begin{equation}\label{E:loglik}
 \ell (\theta)=\ell (\theta;y)=\sum^n_{i=1}\ell_i(\mu_i,\phi),
\end{equation}
where
\begin{align*}
\ell_i(\mu_i,\phi) & = \log \Gamma{\left( \phi \right)}\!- \log \Gamma{\left( \mu_i \phi \right)}\!- \log \Gamma{\left((1-\mu_i)\phi \right)} + \left(\mu_i \phi \!-1 \right) \log y_i \\
&  + \left\{(1-\mu_i)\phi\!-1 \right\}\log(1-y_i). 
\end{align*}

The score function $U(\theta)$ is obtained by differentiating the log-likelihood function with respect to unknown parameters. The score function with respect to $\beta$ and $\phi$ are, respectively,
\begin{align*}
& U_\beta(\theta)=\phi X^{\top}T(y^{\ast}-\mu^{\ast}),\\
& U_\phi(\theta)=\sum^n_{i=1}{\mu_i(y^{\ast}_i-\mu^{\ast})+\log(1-y_i)-\psi((1-\mu_i)\phi)+\psi(\phi)},
\end{align*}
where $X$ is the $n\times p$ covariates matrix whose $i$-th row is $x^{\top}_i$. Also, $T = \text{diag}\{1/g^{\prime}(\mu_1),\ldots,1/g^{\prime}(\mu_n)\}$, $y^{\ast}=\{y^{\ast}_1,\ldots,y^{\ast}_n\}^{\top}$, $\mu^{\ast}=\{\mu^{\ast}_1,\ldots,\mu^{\ast}_n\}^{\top}$,
$y^{\ast}_i \! = \log \left( \frac{y_i}{1-y_i} \right)$, $\mu^{\ast}_i=\psi(\mu_i\phi)-\psi((1-\mu_i)\phi)$ and $\psi(\cdot)$ is the digamma function\footnote{The polygamma function is defined, for $m=0,1,\ldots$, as $\psi^{(m)}(x)=\left({\rm d}^{m+1}/{\rm d}x^{m+1} \right)\log\Gamma(x)$, $x>0$. The digamma function is obtained by setting $m=0$.}.

The maximum likelihood estimators are the solution to the following system:
\begin{equation*}
\left\{ \begin{array}{ll}
U_\beta(\theta)= &0 \\
U_\phi(\theta)= &0
\end{array} \right. .
\end{equation*}
The maximum likelihood estimators, $\hat{\beta}$ and $\hat{\phi}$, cannot be expressed in closed-form. They are typically obtained by numerically maximizing the log-likeli\-hood function using a Newton or quasi-Newtion nonlinear optimization algorithm. For details on nonlinear optimization algorithms, see \cite{Press1992}.

Fisher's joint information for $\beta$ and $\phi$ is given by
\begin{equation*}
K=K(\theta) = \left(
\begin{array}{cc}
	K_{(\beta,\beta)} & K_{(\beta,\phi)} \\
	K_{(\phi,\beta)} & K_{(\phi,\phi)} 
\end{array}
 \right),
\end{equation*}
where $K_{(\beta,\beta)} = \phi X^{\top} WX $, $K_{(\beta,\phi)} = (K_{(\phi,\beta)})^{\top}=X^{\top}Tc$ and $K_{(\phi,\phi)}=\mathrm{tr}(D)$. 
Here, $W$ ($n\times n$ diagonal matrix), $c$ ($n$-vector) and $D$ ($n\times n$ diagonal matrix) have typical elements given by 
$
w_i = \phi \left\{ \psi^{\prime}(\mu_i \phi)+\psi^{\prime}((1-\mu_i)\phi)\right\}\frac{1}{\{g^{\prime}(\mu_i)\}^2}
$,
$
\,c_i = \phi \left\{ \psi^{\prime}(\mu_i \phi)\mu_i -\psi^{\prime}((1-\mu_i)\phi)(1-\mu_i)\right\}
$,
$
\,d_i = \psi^{\prime}(\mu_i \phi)\mu_i^2 +\psi^{\prime}((1-\mu_i)\phi)(1-\mu_i)^2-\psi^{\prime}(\phi)
$, respectively. That is,  
$W = \text{diag}\{w_1,\ldots,w_n\}$, 
$c=(c_1,\ldots,c_n)^{\top}$ and
$D = \text{diag}\{d_1,\ldots,d_n\}$. For details on log-likelihood derivatives, see \ref{A:cumulants}.

Under mild regularity conditions, and in large samples, the joint distribution of $\hat{\beta}$ and $\hat{\phi}$ is approximately $k$-multivariate normal: 
\begin{equation*}
\left(\begin{array}{c}
\hat{\beta} \\
\hat{\phi}
\end{array}\right)\sim
\mathcal{N}_{k}
\left(\left(\begin{array}{c}
{\beta} \\
{\phi}
\end{array}\right),K^{-1}\right),
\end{equation*}
approximately.

\section{Improved likelihood ratio testing inference}\label{S:correction}

Consider the parametric model presented in (\ref{E:g_mu}) and the corresponding log-likelihood function given in (\ref{E:loglik}), where $\theta = (\theta_1^{\top},\theta_2^{\top})^{\top}$ is the model $k$-dimensional parametric vector, $\theta_1$ being a $q$-dimensional vector and $\theta_2$ containing the remaining $k-q$ parameters. Suppose that we wish test the null hypothesis 
$$\mathcal{H}_0\!:\, \theta_1 = \theta_1^{0}$$ 
against the alternative hypothesis
$$\mathcal{H}_1\!:\, \theta_1 \neq \theta_1^{0},$$
where $\theta_1^{0}$ is a given $q\times 1$ vector of scalars. Hence, $\theta_2$ is the vector of nuisance parameters and $\theta_1$ is the vector of parameters of interest. The null hypothesis imposes $q$ restrictions on the parameter vector. 
The likelihood ratio test statistic can be written as
\begin{equation*}
LR = 2\{\ell(\hat{\theta};y)-\ell(\tilde{\theta};y)\},
\end{equation*}
where the vector $\tilde{\theta}$ is the restricted maximum likelihood estimator of $\theta$ obtained by imposing the null hypothesis, i.e., $\tilde{\theta}=({\theta_1^{0}}^{\top},\tilde{\theta_2}^{\top})^{\top}$.

In large samples, the likelihood ratio statistic $LR$ is approximately distributed as $\chi^2_q$ under $\mathcal{H}_0$ with error of the order $n^{-1}$. In small samples, however, this approximation may be poor. Since the test is conducted using critical values obtained from the limiting null distribution ($\chi_q^2$) and that such a distribution may provide a poor approximation to the test statistic exact null distribution in small samples, the likelihood ratio test may be considerably size distorted.

Likelihood ratio testing inference can be made more accurate by applying a correction factor to the test statistic. This correction factor is known as the Bartlett correction and was proposed by \cite{Bartlett1937} and later generalized by \cite{Lawley1956}. The underlying idea is to base inferences on the modified statistic given by $LR/c$, 
where $c={\rm E}(LR)/q$ is the Bartlett correction factor. It is possible to express the Bartlett correction factor $c$ using moments of log-likelihood derivatives; see \cite{Lawley1956}. It is noteworthy that the Bartlett correction delivers an improvement in the rate at which size distortions vanish; see \cite{Barndorff1984}. In particular, $\Pr(LR \leq x)=\Pr(\chi^2_q \leq x) + O(n^{-1})$ and $\Pr(LR/c \leq x)=\Pr(\chi^2_q \leq x) + O(n^{-2})$.

\subsection{A matrix formula for the Bartlett correction factor}

The Bartlett correction factor can be written as 
\begin{equation*}
c=1+\frac{\epsilon_k - \epsilon_{k-q}}{q}.
\end{equation*}
Using Lawley's expansion \citep{Lawley1956}, the expected value of the likelihood ratio statistic can be expressed as 
\begin{equation*}
{\rm E}(LR)=q+\epsilon_k -\epsilon_{k-q}+O(n^{-2}),
\end{equation*}
where
\begin{align}
 \epsilon_k&=\sum_{\theta}(\lambda_{rstu}-\lambda_{rstuvw}), \label{E:soma_ek}\\
 \lambda_{rstu}&=\kappa^{rs}\kappa^{tu}\left\{\frac{\kappa_{rstu}}{4}-\kappa_{rst}^{(u)}+\kappa_{rt}^{(su)}\right\}, \nonumber \\
 \lambda_{rstuvw}&=\kappa^{rs}\kappa^{tu}\kappa^{vw}\left\{
\kappa_{rtv} \left(\frac{\kappa_{suw}}{6}-\kappa_{sw}^{(u)} \right)+
\kappa_{rtu} \left(\frac{\kappa_{svw}}{4}-\kappa_{sw}^{(v)} \right)+
\kappa_{rt}^{(v)}\kappa_{sw}^{(u)}\right. \nonumber \\ 
 &\left. +\kappa_{rt}^{(u)}\kappa_{sw}^{(v)} \right\} \nonumber
\end{align}
and
\begin{equation*}
\kappa_{rs}={\rm E}\left(\frac{\partial^2\ell(\theta)}{\partial \theta_r \partial \theta_s}\right), \; \kappa_{rst}={\rm E}\left(\frac{\partial^3\ell(\theta)}{\partial \theta_r \partial \theta_s \partial \theta_t}\right), \; \kappa_{rs}^{(t)}=\frac{\partial \kappa_{rs}}
{\theta_t}, \; \mbox{etc}.
\end{equation*}
Notice that $-\kappa^{rs}$ is the $(r,s)$ element of the inverse of Fisher's information matrix, $K^{-1}$. The summation in (\ref{E:soma_ek}) runs over all components of $\theta$, i.e., the indices $r$, $s$, $t$, $u$, $v$ and $w$ vary over all $k$ parameters. The expression for $\epsilon_{k-q}$ is obtained from (\ref{E:soma_ek}) by letting summation to only run over the nuisance parameters in $\theta_2$. All $\kappa$'s are of order $n$, and $\epsilon_{k}$ and $\epsilon_{k-q}$ are of order $O(n^{-1})$.

It can be quite hard to derive the Bartlett correction using Lawley's general formula, since it involves the product of mixed cumulants that are not invariant under index permutations \citep{Cordeiro1993}. In particular, in the beta regression model  the parameters $\beta$ and $\phi$ are not orthogonal, i.e., Fisher's information matrix is not block diagonal, and as consequence the Bartlett correction derivation via the Lawley's approach becomes especially cumbersome. An alternative is to use the general matrix formula given by \cite{Cordeiro1993}.

In order to express $\epsilon_k$ in matrix form, we first define the following $k \times k$ matrices: $A^{(tu)}$, $P^{(t)}$ and $Q^{(u)}$, for $t,u=1,\ldots,k$. The $(r,s)$ elements of such matrices are
\begin{equation}
A^{(tu)}= \left\{ \frac{\kappa_{rstu}}{4}-\kappa_{rst}^{(u)}+\kappa_{rt}^{(su)}\right\}, \quad 
P^{(t)}=\{\kappa_{rst}\}, \quad
Q^{(u)}=\{\kappa_{su}^{(r)}\}, 
\end{equation}
for $t,u=1,\ldots,k$. Using matrix notation, we can write
\begin{align}
& \sum_{\theta}\lambda_{rstu}={\rm tr}(K^{-1}L), \nonumber\\
& \sum_{\theta}\kappa^{rs}\kappa^{tu}\kappa^{vw}\left\{
 \frac{1}{6}\kappa_{rtv}\kappa_{suw}-\kappa_{rtv}\kappa_{sw}^{(u)}+
\kappa_{rt}^{(v)}\kappa_{sw}^{(u)}
 \right\} \nonumber \\
& \quad = -\frac{1}{6}{\rm tr}(K^{-1}M_1)+{\rm tr}(K^{-1}M_2)-{\rm tr}(K^{-1}M_3), \\
& \sum_{\theta}\kappa^{rs}\kappa^{tu}\kappa^{vw}\left\{
 \frac{1}{4}\kappa_{rtu}\kappa_{svw}-\kappa_{rtu}\kappa_{sw}^{(v)}+
\kappa_{rt}^{(u)}\kappa_{sw}^{(v)}
 \right\} \nonumber\\
& \quad = -\frac{1}{4}{\rm tr}(K^{-1}N_1)+{\rm tr}(K^{-1}N_2)-{\rm tr}(K^{-1}N_3)\label{E:last_bartlett},
\end{align} 
where the $(r,s)$ elements of the $L$, $M_1$, $M_2$, $M_3$, $N_1$, $N_2$ and $N_3$ matrices are given, respectively, by
\begin{align*}
& \left\{{\rm tr}(K^{-1}A^{(rs)})\right\}, \\
& \left\{{\rm tr}(K^{-1}P^{(r)}K^{-1}P^{(s)})\right\}, \\
& \left\{{\rm tr}(K^{-1}P^{(r)}K^{-1}{Q^{(s)}}^{\top})\right\}, \\
& \left\{{\rm tr}(K^{-1}Q^{(r)}K^{-1}Q^{(s)})\right\}, \\
& \left\{{\rm tr}(P^{(r)}K^{-1}){\rm tr}(P^{(s)}K^{-1})\right\}, \\
& \left\{{\rm tr}(P^{(r)}K^{-1}){\rm tr}(Q^{(s)}K^{-1})\right\}, \\
& \left\{{\rm tr}(Q^{(r)}K^{-1}){\rm tr}(Q^{(s)}K^{-1})\right\}.  \\
\end{align*}

Using (\ref{E:soma_ek})--(\ref{E:last_bartlett}) we can write 
\begin{equation*}\label{E:epsilon_k}
\epsilon_k = {\rm tr}\left[K^{-1}(L-M-N)\right],
\end{equation*}
where $M=-\frac{1}{6}M_1+M_2-M_3$ and $N=-\frac{1}{4}N_1+N_2-N_3$.

The term in (\ref{E:epsilon_k}) can be easily computed using a matrix programming language, like \texttt{Ox} \citep{Doornik2007} and {\tt R} \citep{R2009}. It only requires the computation of $(k+1)^2$ matrices of order $k$, namely:  $K^{-1}$, $k$ matrices $P^{(t)}$, $k$ matrices $Q^{(u)}$ and $k^2$ matrices $A^{(tu)}$. The remaining matrices can be obtained from them using simple matrix operations. Thus, to obtain the Bartlett correction factor $c$ we need compute $(k+1)^2$ matrices of order $k$ and $(k-q+1)^2$ matrices of order $k-q$. In order to obtain the matrices $P^{(t)}$, $Q^{(u)}$ and $A^{(tu)}$ we need cumulants of log-likelihood derivatives up to fourth order. We have derived these cumulants for the beta regression model and present them in \ref{A:cumulants}. 
 
The usual Bartlett corrected likelihood ratio statistic is given by $LR/c$. There are, however, two other equivalent specifications that deliver the same order of accuracy. The three Bartlett corrected test statistics are
\begin{align*}
& LR_{b1} = \frac{LR}{c},\\
& LR_{b2} = LR \exp \left\{-\frac{(\epsilon_k -\epsilon_{k-q})}{q} \right\},\\
& LR_{b3} = LR \left\{ 1- \frac{(\epsilon_k -\epsilon_{k-q})}{q} \right\}.
\end{align*}
The corrected statistics $LR_{b1}$, $LR_{b2}$ and $LR_{b3}$ are equivalent to order $O(n^{-1})$ \citep{Lemonte2010}, and $LR_{b2}$ has the advantage of only taking positive values.

\subsection{Bootstrap Bartlett correction} 

\cite{Rocke1989} introduced a numeric alternative to the analytic Bartlett correction in which the correction factor is computed using Efron's bootstrap \citep{Efron1979}. The bootstrap Bartlett correction can be described as follow. Bootstrap resamples are used to estimate the likelihood ratio statistic expected value. Here, $B$ bootstrap resamples ($y^{*1},y^{*2},\ldots,y^{*B}$) are generated using the parametric bootstrap and imposing $\mathcal{H}_0$. Data generation is performed from the postulated model after replacing the unknown parameter vector by its restricted estimate, i.e., by the estimate obtained under the null hypothesis. For each pseudo sample $y^{*b}$, $b=1,2,\ldots,B$, the $LR$ statistic is computed as
\begin{equation*}
LR^{*b} = 2\{\ell(\hat{\theta}^{*b};y^{*b}) - \ell(\tilde{\theta}^{*b};y^{*b})\},
\end{equation*}
where $\hat{\theta}^{*b}$ and $\tilde{\theta}^{*b}$ are the maximum likelihood estimators of $\theta$ obtained from the maximization of $\ell(\theta;y^{*b})$ under $\mathcal{H}_1$ and $\mathcal{H}_0$, respectively. The bootstrap Bartlett corrected likelihood ratio statistic is then computed as 
\begin{equation*}
LR_{boot} = \frac{LR \, q}{\overline{LR^*}}, 
\end{equation*} 
where $\overline{LR^*}={B}^{-1}\sum_{b=1}^B{LR^{*b}}$.

It is noteworthy that the bootstrap Bartlett correction is computationally more efficient than the usual approach of using the bootstrap method to obtain a critical value (or a $p$-value) since it requires a smaller number of resamples. The usual bootstrap approach typically requires 1,000 bootstrap resamples, since it involves estimating tail quantities \citep{Efron1986, Efron1987}; on the other hand, the bootstrap Bartlett correction is expected to work well when based on only 200 artificial samples. Notice that in the latter we use data resampling to estimate the mean of a distribution, and not an upper quantile. According to \cite{Rocke1989} the bootstrap Bartlett correction that uses $B=100$ typically yields inferences that are as accurate as those obtained using the usual bootstrapping scheme with $B=700$.

\subsection{Skovgaard's adjustment} 

In a different approach, \cite{Skovgaard2001} generalized the results in \cite{Skovgaard1996} and presented a much simpler way to improve likelihood ratio testing inference. His adjustment was later computed for various models; see, e.g., \cite{Ferrari2008}, \cite{Pinheiro2011}, \cite{Melo2009} and \cite{Pereira2010}. The numerical evidence presented by these authors indicates that hypothesis testing inference based on Skovgaard's modified likelihood ratio statistic is typically more accurate than that based on the uncorrected statistic.

In order to present the Skovgaard's adjustment to the likelihood ratio test statistic, which was derived by \cite{Pinheiro2011} for beta regressions, we shall now introduce some additional notation. Recall that $\theta = (\theta_1^{\top},\theta_2^{\top})^{\top}$, where $\theta_1$ and $\theta_2$ are the interest and nuisance parameters, respectively. Let $J$ denote the observed information matrix and let $J_{1 1}$ be the observed information matrix corresponding to $\theta_1$. Additionally, $\hat{J}=J(\hat{\theta})$, $\tilde{J}=J(\tilde{\theta})$, $\hat{K}=K(\hat{\theta})$, $\tilde{K}=K(\tilde{\theta})$ and $\tilde{U}=U(\tilde{\theta})$.

The \citeauthor{Skovgaard2001} modified likelihood ratio test statistic is given by
\begin{equation*}
LR_{sk1} = LR - 2\log \xi, 
\end{equation*} 
where
\begin{equation*}
\xi = \frac{
\{|\tilde{K}|\,|\hat{K}|\,|\tilde{J}_{11}|\}^{1/2}   
}
{
|\bar{\Upsilon}|\,|\{\tilde{K}\bar{\Upsilon}^{-1}\hat{J}\,\hat{K}^{-1}\,\bar{\Upsilon}\}_{11}|^{1/2}
} \frac{
\{\tilde{U}^{\top}\bar{\Upsilon}^{-1}\hat{K}\,\hat{J}^{-1}\bar{\Upsilon}\,\tilde{K}^{-1}\tilde{U}\}^{q/2}
}
{
LR^{q/2-1}\tilde{U}^{\top}\bar{\Upsilon}^{-1}\bar{\upsilon}
}.
\end{equation*} 
Here, $\bar{\Upsilon}$ and $\bar{\upsilon}$ are obtained by 
replacing $\theta$ for $\hat{\theta}$ and $\theta_2$ for $\tilde{\theta}$ in 
$
\Upsilon={\rm E}_{\theta}[U(\theta)U^{\top}(\theta_2)]
$
and
$
\upsilon={\rm E}_{\theta}[U(\theta)(\ell(\theta)-\ell(\theta_2))]
$ 
after expected values are computed.

An asymptotically equivalent version of the above test statistic is 
\begin{equation*}
LR_{sk2} = LR\left(1-\frac{1}{LR}\log \xi\right)^2.
\end{equation*} 

Under $\mathcal{H}_0$, $LR_{sk1}$ and $LR_{sk2}$ are approximately distributed as $\chi^2_q$ with a high degree of accuracy \citep{Skovgaard2001, Pinheiro2011}. For more details and matrix formulas for $\bar{\Upsilon}$ and $\bar{\upsilon}$ in the beta regressions, see \cite{Pinheiro2011}. In \cite{Pinheiro2011} the Skovgaard adjustment is derived for a general class of beta regressions that allows for nonlinearities and varying dispersion.

\section{Numerical evidence}\label{S:simulation}

This section presents Monte Carlo simulation results on the small sample performance of the likelihood ratio test ($LR$) in beta regression and also of six tests that are based on corrected statistics, namely: the three Bartlett corrected statistics ($LR_{b1}$, $LR_{b2}$ and $LR_{b3}$), the bootstrap Bartlett corrected statistic ($LR_{boot}$) and the two modified statistics obtained using Skovgaard's adjustment ($LR_{sk1}$ and $LR_{sk2}$). The number of Monte Carlo replications is 10,000. For each Monte Carlo replication we performed 500 bootstrap replications. All simulations were carried out using the {\tt R} programming language \citep{R2009}.

We consider the following beta regression the model:
\begin{equation*}
{\rm logit}(\mu_i) = \beta_1 +\beta_2 x_{2i} +\beta_3 x_{3i} +\beta_4 x_{4i} +\beta_5 x_{5i}.
\end{equation*}
The covariates values are chosen as random draws from the $\mathcal{U}(-0.5,0.5)$ distribution and are kept fixed during the simulations. We use four different values for the precision parameter $\phi$, namely: $100$, $30$, $10$ and $5$. Restrictions on $\beta$ are tested using samples of 15, 20, 30 and 40 observations and at three nominal levels: $\alpha=10\%$, $5\%$ and $1\%$. The null hypotheses are $\mathcal{H}_0:\,\beta_2=0$ ($q=1$), $\mathcal{H}_0:\,\beta_2=\beta_3=0$ ($q=2$) and $\mathcal{H}_0:\,\beta_2=\beta_3=\beta_4=0$ ($q=3$), to be tested against two-sided alternative hypotheses. When $q=1$, we set $\beta_1=1$, $\beta_2=0$, $\beta_3=1$, $\beta_4=5$ and $\beta_5=-4$. When $q=2$, $\beta_1=1$, $\beta_2=\beta_3=0$, $\beta_4=5$ and $\beta_5=-4$. Finally, when $q=3$, the parameter values used for data generation are $\beta_1=1$, $\beta_2=\beta_3=\beta_4=0$ and $\beta_5=-4$. 

Tables \ref{T:size1} ($q=1$), \ref{T:size2} ($q=2$)  and \ref{T:size3} ($q=3$) present the null rejection rates of the different tests. The figures in these tables clearly show that the likelihood ratio test is considerably oversized (liberal); its null rejection rate can be eight times larger than the nominal level, as in Table \ref{T:size2} for $\phi=5$, $\alpha=1\%$ and $n=15$. In general, larger sample sizes and/or larger values of $\phi$ lead to smaller size distortions.

\begin{table}[t]
\caption{Null rejection rates (\%) for the test of $\mathcal{H}_0:\,\beta_2=0$ ($q=1$).} \label{T:size1}
\tablesize
\begin{center}
\begin{tabular}{ll|cccc|cccc|cccc}
\hline
 & & \multicolumn{4}{|c|}{$\alpha = 10\%$} & \multicolumn{4}{c|}{$\alpha = 5\%$} & \multicolumn{4}{c}{$\alpha = 1\%$} \\
\hline 		
\!\!$\phi$	& \backslashbox[10pt][c]{Stat}{\!\!$n$} & 15 & 20 & 30 & 40 & 15 & 20 & 30 & 40 & 15 & 20 & 30 & 40 \\
\hline
	&	$LR$		&	18.9	&	16.5	&	13.7	&	12.8	&	11.7&	9.5	&	7.5	&	7.2	&	4.0	&	3.0	&	2.1	&	1.7	\\
	&	$LR_{b1}$	&	12.4	&	11.6	&	10.6	&	10.5	&	6.9	&	5.9	&	5.4	&	5.7	&	1.6	&	1.4	&	1.0	&	1.0	\\
	&	$LR_{b2}$	&	11.5	&	11.0	&	10.3	&	10.3	&	6.2	&	5.5	&	5.3	&	5.6	&	1.4	&	1.2	&	1.0	&	1.0	\\
\!\!100	\!\!\!\!\!\!&$LR_{b3}$&	10.0&	10.0&	9.9		&	10.1	&	5.0	&	4.9	&	5.1	&	5.4	&	0.9	&	1.0	&	0.9	&	1.0	\\
	&	$LR_{sk1}$	&	10.1	&	10.0	&	9.9		&	10.3	&	4.9	&	4.9	&	5.2	&	5.4	&	1.0	&	1.0	&	1.0	&	1.0	\\
	&	$LR_{sk2}$	&	11.8	&	11.4	&	11.0	&	11.2	&	6.3	&	5.9	&	6.1	&	6.3	&	1.8	&	1.6	&	1.6	&	1.5	\\
	&	$LR_{boot}$	&	10.2	&	10.1	&	9.9		&	10.4	&	5.1	&	5.0	&	5.1	&	5.5	&	0.9	&	1.0	&	1.0	&	1.0	\\
	
\hline
	&	$LR$	&	19.5	&	16.8	&	14.8	&	13.0	&	12.1	&	9.7	&	8.0	&	7.0	&	4.2	&	2.7	&	2.3	&	1.6	\\
	&	$LR_{b1}$	&	12.7	&	11.8	&	11.3	&	10.5	&	6.8	&	6.1	&	6.1	&	5.2	&	1.7	&	1.3	&	1.4	&	1.1	\\
	&	$LR_{b2}$	&	11.7	&	11.2	&	11.0	&	10.2	&	6.1	&	5.6	&	6.0	&	5.1	&	1.4	&	1.1	&	1.2	&	1.1	\\
\!\!30	&	$LR_{b3}$	&	10.2	&	10.3	&	10.6	&	10.0	&	5.1	&	5.0	&	5.7	&	4.9	&	1.1	&	1.0	&	1.1	&	1.0	\\
	&	$LR_{sk1}$	&	10.2	&	10.3	&	10.8	&	10.2	&	5.2	&	4.9	&	5.7	&	5.0	&	1.2	&	1.0	&	1.2	&	1.0	\\
	&	$LR_{sk2}$	&	13.2	&	11.7	&	12.7	&	11.7	&	7.6	&	6.2	&	7.2	&	6.2	&	2.7	&	1.7	&	2.1	&	2.0	\\
	&	$LR_{boot}$	&	10.2	&	10.2	&	10.6	&	10.3	&	4.9	&	5.0	&	5.6	&	4.9	&	1.1	&	1.0	&	1.2	&	1.1	\\
\hline
	&	$LR$	&	22.0	&	21.4	&	17.9	&	13.7	&	14.4	&	13.8	&	11.0	&	8.1	&	5.5	&	5.1	&	3.6	&	2.2	\\
	&	$LR_{b1}$	&	15.2	&	15.9	&	14.2	&	11.4	&	8.6	&	9.1	&	8.2	&	6.3	&	2.4	&	2.5	&	2.2	&	1.4	\\
	&	$LR_{b2}$	&	13.8	&	15.1	&	13.9	&	11.2	&	7.7	&	8.5	&	7.9	&	6.2	&	1.9	&	2.2	&	2.0	&	1.3	\\
\!\!10	&	$LR_{b3}$	&	12.0	&	14.2	&	13.5	&	11.0	&	6.3	&	7.8	&	7.6	&	6.0	&	1.5	&	1.9	&	1.9	&	1.2	\\
	&	$LR_{sk1}$	&	12.1	&	14.6	&	13.9	&	11.0	&	6.4	&	8.0	&	7.8	&	5.9	&	1.5	&	2.0	&	2.0	&	1.2	\\
	&	$LR_{sk2}$	&	14.9	&	17.3	&	16.3	&	13.0	&	8.7	&	10.2	&	9.9	&	7.6	&	2.8	&	3.5	&	3.6	&	2.6	\\
	&	$LR_{boot}$	&	12.2	&	14.6	&	14.4	&	12.7	&	6.6	&	8.2	&	8.5	&	7.2	&	1.5	&	2.1	&	2.2	&	1.9	\\
\hline
	&	$LR$	&	19.1	&	16.2	&	15.4	&	12.7	&	12.2	&	9.6	&	8.7	&	6.8	&	4.3	&	3.0	&	2.5	&	1.8	\\
	&	$LR_{b1}$	&	12.9	&	11.5	&	12.0	&	10.8	&	7.0	&	6.2	&	6.3	&	5.3	&	1.8	&	1.3	&	1.3	&	1.2	\\
	&	$LR_{b2}$	&	12.1	&	11.0	&	11.6	&	10.7	&	6.3	&	5.8	&	6.0	&	5.2	&	1.4	&	1.2	&	1.3	&	1.1	\\
\!\!5	&	$LR_{b3}$	&	10.6	&	10.2	&	11.3	&	10.5	&	5.3	&	5.2	&	5.8	&	5.1	&	0.9	&	1.0	&	1.2	&	1.1	\\
	&	$LR_{sk1}$	&	11.7	&	10.9	&	11.5	&	10.7	&	6.2	&	5.6	&	6.1	&	5.3	&	1.3	&	1.1	&	1.2	&	1.2	\\
	&	$LR_{sk2}$	&	15.2	&	14.1	&	14.1	&	13.1	&	9.1	&	8.1	&	8.2	&	7.3	&	3.4	&	2.8	&	2.8	&	2.8	\\
	&	$LR_{boot}$	&	13.9	&	10.4	&	11.7	&	12.3	&	7.8	&	5.5	&	6.2	&	6.6	&	2.5	&	1.1	&	1.4	&	1.8	\\
\hline
\end{tabular}
\end{center}
\end{table}

\begin{table}[t]
\caption{Null rejection rates (\%) for the test of $\mathcal{H}_0:\,\beta_2=\beta_3=0$ ($q=2$).} \label{T:size2}
{
\tablesize
\begin{center}
\begin{tabular}{ll|cccc|cccc|cccc}
\hline
 & & \multicolumn{4}{|c|}{$\alpha = 10\%$} & \multicolumn{4}{c|}{$\alpha = 5\%$} & \multicolumn{4}{c}{$\alpha = 1\%$} \\
\hline 			
\!\!$\phi$	& \backslashbox[10pt][c]{Stat}{\!\!$n$} & 15 & 20 & 30 & 40 & 15 & 20 & 30 & 40 & 15 & 20 & 30 & 40 \\
\hline
	&	$LR$		&	22.0	&	17.1	&	14.1&	13.7&	14.1&	10.0&	7.8	&	7.5	&	4.9	&	3.1	&	2.2	&	1.7	\\
	&	$LR_{b1}$	&	13.3	&	10.9	&	10.1&	10.5&	7.3	&	5.9	&	5.1	&	5.6	&	1.6	&	1.3	&	1.2	&	1.1	\\
	&	$LR_{b2}$	&	12.1	&	10.3	&	9.7	&	10.3&	6.4	&	5.5	&	5.0	&	5.5	&	1.3	&	1.2	&	1.2	&	1.0	\\
\!\!100\!\!\!\!\!\!&$LR_{b3}$&10.3&	9.5		&	9.4	&	10.1&	5.4	&	4.8	&	4.7	&	5.4	&	1.0	&	1.0	&	1.1	&	1.0	\\
	&	$LR_{sk1}$	&	10.4	&	9.6		&	9.5	&	10.1&	5.3	&	4.9	&	4.7	&	5.4	&	1.0	&	1.0	&	1.1	&	1.0	\\
	&	$LR_{sk2}$	&	11.5	&	10.1	&	9.7	&	10.3&	6.1	&	5.2	&	4.9	&	5.5	&	1.2	&	1.1	&	1.2	&	1.0	\\
	&	$LR_{boot}$	&	10.5	&	9.6		&	9.5	&	10.2&	5.4	&	5.0	&	4.8	&	5.4	&	1.0	&	1.0	&	1.1	&	1.0	\\
\hline
	&	$LR$	&	23.0	&	17.8	&	14.6	&	13.8	&	14.4	&	10.6	&	7.8	&	7.6	&	5.4	&	3.3	&	1.9	&	1.9	\\
	&	$LR_{b1}$	&	13.7	&	11.7	&	10.2	&	10.7	&	7.8	&	6.0	&	4.8	&	5.4	&	2.0	&	1.4	&	1.0	&	1.0	\\
	&	$LR_{b2}$	&	12.4	&	10.9	&	9.8	&	10.5	&	6.9	&	5.6	&	4.7	&	5.3	&	1.5	&	1.2	&	1.0	&	1.0	\\
\!\!30	&	$LR_{b3}$	&	10.7	&	10.1	&	9.4	&	10.2	&	5.6	&	5.0	&	4.5	&	5.2	&	1.1	&	1.0	&	1.0	&	0.9	\\
	&	$LR_{sk1}$	&	11.2	&	10.3	&	9.6	&	10.3	&	6.4	&	5.1	&	4.5	&	5.3	&	1.8	&	1.0	&	1.0	&	0.9	\\
	&	$LR_{sk2}$	&	12.2	&	10.9	&	9.9	&	10.5	&	7.1	&	5.4	&	4.7	&	5.4	&	1.9	&	1.2	&	1.1	&	1.0	\\
	&	$LR_{boot}$	&	10.5	&	10.1	&	9.5	&	10.4	&	5.6	&	5.0	&	4.6	&	5.2	&	1.1	&	1.0	&	1.0	&	1.0	\\
\hline
	&	$LR$	&	26.0	&	19.1	&	16.0	&	15.2	&	17.4	&	11.8	&	9.1	&	8.4	&	7.0	&	3.7	&	2.7	&	2.4	\\
	&	$LR_{b1}$	&	16.5	&	12.7	&	11.7	&	12.0	&	9.8	&	6.7	&	6.3	&	6.2	&	2.8	&	1.6	&	1.4	&	1.4	\\
	&	$LR_{b2}$	&	15.1	&	12.0	&	11.3	&	11.8	&	8.9	&	6.3	&	6.0	&	6.0	&	2.3	&	1.4	&	1.3	&	1.4	\\
\!\!10	&	$LR_{b3}$	&	13.2	&	11.0	&	10.9	&	11.6	&	7.4	&	5.7	&	5.6	&	5.9	&	1.8	&	1.3	&	1.2	&	1.3	\\
	&	$LR_{sk1}$	&	13.4	&	11.5	&	11.0	&	11.7	&	7.5	&	5.9	&	5.7	&	6.0	&	1.8	&	1.3	&	1.3	&	1.3	\\
	&	$LR_{sk2}$	&	14.5	&	12.2	&	11.4	&	12.1	&	8.4	&	6.4	&	6.0	&	6.3	&	2.2	&	1.5	&	1.4	&	1.5	\\
	&	$LR_{boot}$	&	13.6	&	11.0	&	11.1	&	12.8	&	7.8	&	5.6	&	5.8	&	6.8	&	2.0	&	1.2	&	1.3	&	1.7	\\
\hline
	&	$LR$	&	27.8	&	19.7	&	15.3	&	13.1	&	19.3	&	12.0	&	8.5	&	7.0	&	8.0	&	4.2	&	2.4	&	1.9	\\
	&	$LR_{b1}$	&	18.6	&	13.1	&	11.2	&	10.1	&	11.0	&	7.1	&	5.8	&	5.5	&	3.6	&	1.8	&	1.2	&	1.2	\\
	&	$LR_{b2}$	&	17.2	&	12.4	&	10.8	&	10.0	&	10.0	&	6.5	&	5.6	&	5.4	&	3.1	&	1.7	&	1.1	&	1.1	\\
\!\!5	&	$LR_{b3}$	&	14.9	&	11.5	&	10.5	&	9.8	&	8.4	&	6.0	&	5.4	&	5.2	&	2.3	&	1.5	&	1.0	&	1.0	\\
	&	$LR_{sk1}$	&	14.4	&	12.0	&	11.2	&	10.0	&	7.9	&	6.2	&	5.6	&	5.2	&	2.2	&	1.6	&	1.1	&	1.2	\\
	&	$LR_{sk2}$	&	16.0	&	12.8	&	11.5	&	10.4	&	9.1	&	6.7	&	5.9	&	5.6	&	2.7	&	1.8	&	1.2	&	1.3	\\
	&	$LR_{boot}$	&	15.4	&	12.1	&	11.0	&	14.8	&	8.9	&	6.4	&	5.8	&	8.7	&	2.6	&	1.7	&	1.2	&	2.5	\\
\hline
\end{tabular}
\end{center}}
\end{table}

The simulation results for $q=1$ presented in Table \ref{T:size1} indicate that the corrected tests display good small sample behavior. The Bartlett corrected test $LR_{b3}$ is the best performer, being followed by the Skovgaard adjusted test $LR_{sk1}$ and by the bootstrap Bartlett corrected test, $LR_{boot}$. The latter outperforms the competition when $\phi=30$. For instance, when $\phi=30$ and $\alpha=10\%$, the null rejection rates of  $LR_{b3}$ for the four sample sizes are 10.2\%, 10.3\%, 10.6\% and 10.0\% and the corresponding rates of the $LR_{sk1}$ are 10.2\%, 10.3\%, 10.8\% and 10.2\%. The good performance of the $LR_{b3}$ test can be observed in all scenarios.

\begin{table}[t]
\caption{Null rejection rates (\%) for the test of $\mathcal{H}_0:\,\beta_2=\beta_3=\beta_4=0$ ($q=3$).} \label{T:size3}
{
\tablesize
\begin{center}
\begin{tabular}{ll|cccc|cccc|cccc}
\hline
 & & \multicolumn{4}{|c|}{$\alpha = 10\%$} & \multicolumn{4}{c|}{$\alpha = 5\%$} & \multicolumn{4}{c}{$\alpha = 1\%$} \\
\hline 			
\!\!$\phi$	& \backslashbox[10pt][c]{Stat}{\!\!$n$} & 15 & 20 & 30 & 40 & 15 & 20 & 30 & 40 & 15 & 20 & 30 & 40 \\
\hline
	&	$LR$		&	22.3	&	18.5	&	15.5	&	14.0	&	14.4	&	11.1&	8.7	&	7.9	&	5.1	&	3.0	&	2.4	&	2.0	\\
	&	$LR_{b1}$	&	13.2	&	11.7	&	11.2	&	11.0	&	7.4	&	5.8	&	5.5	&	5.4	&	1.8	&	1.3	&	1.2	&	1.1	\\
	&	$LR_{b2}$	&	12.0	&	11.0	&	11.0	&	11.0	&	6.7	&	5.3	&	5.4	&	5.4	&	1.5	&	1.2	&	1.2	&	1.1	\\
\!\!100	\!\!\!\!\!\!
    &	$LR_{b3}$	&	10.4	&	10.2	&	10.5	&	10.7	&	5.4	&	4.8	&	5.1	&	5.3	&	1.1	&	1.1	&	1.1	&	1.1	\\
	&	$LR_{sk1}$	&	10.3	&	10.2	&	10.5	&	10.7	&	5.4	&	4.8	&	5.2	&	5.2	&	1.1	&	1.0	&	1.1	&	1.0	\\
	&	$LR_{sk2}$	&	11.2	&	10.7	&	10.8	&	10.9	&	6.2	&	5.1	&	5.3	&	5.4	&	1.3	&	1.2	&	1.1	&	1.1	\\
	&	$LR_{boot}$	&	10.2	&	10.1	&	10.6	&	10.7	&	5.4	&	4.7	&	5.2	&	5.3	&	1.1	&	1.1	&	1.1	&	1.1	\\
\hline
	&	$LR$	&	23.0	&	17.4	&	14.6	&	13.6	&	14.6	&	10.7	&	8.2	&	7.4	&	4.8	&	3.3	&	2.4	&	1.8	\\
	&	$LR_{b1}$	&	13.1	&	11.2	&	10.3	&	10.4	&	7.0	&	5.9	&	5.5	&	5.2	&	1.8	&	1.3	&	1.1	&	1.2	\\
	&	$LR_{b2}$	&	11.9	&	10.6	&	10.0	&	10.2	&	6.1	&	5.3	&	5.3	&	5.0	&	1.4	&	1.2	&	1.1	&	1.1	\\
\!\!30	&	$LR_{b3}$	&	10.3	&	9.8	&	9.6	&	10.0	&	5.0	&	4.8	&	5.1	&	5.0	&	1.0	&	1.0	&	1.0	&	1.1	\\
	&	$LR_{sk1}$	&	10.2	&	9.9	&	9.7	&	10.1	&	5.1	&	4.8	&	5.1	&	5.1	&	1.1	&	1.0	&	1.0	&	1.1	\\
	&	$LR_{sk2}$	&	10.2	&	10.3	&	9.9	&	10.3	&	5.7	&	5.1	&	5.2	&	5.2	&	1.3	&	1.1	&	1.0	&	1.2	\\
	&	$LR_{boot}$	&	10.2	&	9.8	&	9.5	&	9.9	&	5.0	&	4.8	&	5.1	&	5.0	&	1.0	&	1.0	&	1.0	&	1.2	\\
\hline
	&	$LR$	&	22.1	&	18.6	&	15.3	&	13.6	&	13.7	&	11.2	&	8.7	&	7.5	&	4.6	&	3.2	&	2.2	&	1.8	\\
	&	$LR_{b1}$	&	12.2	&	11.7	&	10.8	&	10.3	&	6.8	&	6.2	&	5.5	&	5.3	&	1.5	&	1.2	&	1.0	&	1.0	\\
	&	$LR_{b2}$	&	11.2	&	11.2	&	10.5	&	10.1	&	6.0	&	5.7	&	5.3	&	5.2	&	1.3	&	1.1	&	1.0	&	1.0	\\
\!\!10	&	$LR_{b3}$	&	9.8	&	10.2	&	10.1	&	9.9	&	5.0	&	5.1	&	5.1	&	5.0	&	0.9	&	0.9	&	0.9	&	1.0	\\
	&	$LR_{sk1}$	&	10.3	&	10.6	&	10.4	&	10.3	&	5.1	&	5.3	&	5.2	&	5.1	&	1.0	&	0.9	&	1.0	&	1.0	\\
	&	$LR_{sk2}$	&	11.2	&	11.1	&	10.7	&	10.4	&	5.8	&	5.7	&	5.4	&	5.2	&	1.2	&	1.0	&	1.0	&	1.0	\\
	&	$LR_{boot}$	&	9.6	&	10.1	&	10.2	&	10.0	&	4.7	&	5.1	&	5.0	&	5.0	&	0.9	&	0.9	&	1.0	&	1.0	\\
\hline
	&	$LR$	&	21.5	&	18.4	&	15.0	&	12.9	&	13.6	&	11.0	&	8.3	&	7.2	&	4.4	&	3.5	&	2.3	&	1.5	\\
	&	$LR_{b1}$	&	12.5	&	11.6	&	10.6	&	9.9	&	6.5	&	6.0	&	5.4	&	5.1	&	1.5	&	1.4	&	1.2	&	0.8	\\
	&	$LR_{b2}$	&	11.3	&	11.0	&	10.3	&	9.8	&	5.9	&	5.5	&	5.2	&	5.1	&	1.3	&	1.3	&	1.1	&	0.8	\\
\!\!5	&	$LR_{b3}$	&	9.7	&	10.2	&	10.0	&	9.7	&	4.8	&	5.0	&	5.0	&	4.9	&	1.0	&	1.0	&	1.0	&	0.8	\\
	&	$LR_{sk1}$	&	10.1	&	10.8	&	10.5	&	10.0	&	5.1	&	5.4	&	5.4	&	5.1	&	1.0	&	1.1	&	1.2	&	0.8	\\
	&	$LR_{sk2}$	&	11.2	&	11.3	&	10.7	&	10.3	&	5.8	&	5.7	&	5.5	&	5.3	&	1.2	&	1.3	&	1.2	&	0.9	\\
	&	$LR_{boot}$	&	9.6	&	10.2	&	10.0	&	9.6	&	4.8	&	5.0	&	5.0	&	5.1	&	1.1	&	1.1	&	1.0	&	0.8	\\
\hline
\end{tabular}
\end{center}}
\end{table}

The results for the cases where we impose more than one restriction, namely $q=2$ and $q=3$, are presented in Tables \ref{T:size2} and \ref{T:size3} and are similar to those obtained for $q=1$. The modified tests once again displayed small size distortions. For instance, for $q=2$, $\phi=30$ and $\alpha=5\%$ the type I error frequency of the uncorrected likelihood ratio test equals $14.4\%$ for $n=15$ whereas for the corrected tests $LR_{b3}$ and $LR_{boot}$ it equals $5.6\%$. The corresponding rejection rate of the $LR_{sk1}$ was $6.4\%$. For $q=3$, $\phi=30$, $\alpha=5\%$ and $n=15$ the null rejection rates are $14.6\%$ ($LR$), $5.0\%$ ($LR_{b3}$) and $5.0\%$ ($LR_{boot}$). For $\phi=30$ and $\alpha=1\%$, the null rejection rates of the $LR_{b3}$, $LR_{sk1}$ and $LR_{boot}$ tests are very close to $1.0\%$ whereas, for the four samples sizes considered, the likelihood ratio test null rejection rates were $4.8\%$, $3.3\%$, $2.4\%$ and $1.8\%$. 

The numerical results presented in Tables \ref{T:size1}, \ref{T:size2} and \ref{T:size3} show that the corrected tests outperform the uncorrected test in small samples. The best performing corrected tests are the Bartlett corrected test $LR_{b3}$, the bootstrap Bartlett corrected test $LR_{boot}$ and the Skovgaard test, $LR_{sk1}$. The null rejection rates of these tests are closer to the nominal levels than those of the uncorrected test and also relative to the other corrected tests. In particular, the bootstrap Bartlett correction works very well when $\phi=30$ and $\phi=100$. 

Table~\ref{T:measures} presents moments and quantiles of the different test statistics alongside with their asymptotic counterparts for $q=2$, $\phi=30$ and $n=20$. It is noteworthy that the $\chi^2_q$ approximation to the likelihood ratio null distribution is quite poor. For example, the limiting null distribution variance equals 4 whereas the variance of $LR$ exceeds 7. On the other hand, the same approximation works quite well for the (analytically and numerically) Bartlett corrected statistics. The $LR_{b3}$ statistic stands out, being followed by $LR_{boot}$. For instance, the mean and variance of $LR_{b3}$ are, respectively, $1.9993$ and $4.0729$, which are very close to two and four, the $\chi^2_2$ mean and variance. The worst performing corrected statistic is $LR_{sk2}$, especially when we consider its skewness and kurtosis. We also note that the limiting null approximation provided to the exact null distribution of $LR_{sk1}$ is not as accurate as for the Bartlett corrected statistics $LR_{b3}$ and $LR_{boot}$. This fact is evidenced by the measures of variance ($4.2331$), skewness ($2.1816$), kurtosis ($11.5872$) and by the 90th quantile (4.6612), which are considerably different from the respective chi-squared reference values. 

\begin{table}[t]
\caption{Estimated moments and quantiles of the different test statistics; $q=2$, $\phi=30$ and $n=20$.} \label{T:measures}
{
\tablesize

\begin{center}
\begin{tabular}{lccccccc}
\hline		
           &  Mean & Variance & Skewness & Kurtosis & 90th-perc & 95th-perc & 99th-perc \\
\hline  
$\chi^2_q$ & $2.0000$ & $4.0000$ & $2.0000$ & $9.0000$ & $4.6052$ & $5.9915$ & $9.2103$ \\
$LR$         & $2.6741$ & $7.2829$ & $2.0784$ & $9.7003$ & $6.1775$ & $8.0134$ & $12.2788$ \\
$LR_{b1}$       & $2.1353$ & $4.6449$ & $2.0799$ & $9.7146$ & $4.9319$ & $6.4065$ & $9.7992$ \\
$LR_{b2}$       & $2.0777$ & $4.3982$ & $2.0804$ & $9.7182$ & $4.7979$ & $6.2333$ & $9.5338$ \\
$LR_{b3}$       & $1.9993$ & $4.0729$ & $2.0810$ & $9.7243$ & $4.6147$ & $5.9960$ & $9.1731$ \\
$LR_{sk1}$      & $2.0127$ & $4.2331$ & $2.1816$ & $11.5872$ & $4.6612$ & $6.0227$ & $9.2845$ \\
$LR_{sk2}$      & $2.0906$ & $4.6776$ & $3.1049$ & $27.7738$ & $4.7836$ & $6.2003$ & $9.5926$ \\
$LR_{boot}$     & $2.0024$ & $4.1168$ & $2.1086$ & $9.9347$ & $4.6103$ & $5.9856$ & $9.2791$ \\
\hline
\end{tabular}
\end{center}}
\end{table}

\begin{figure}[]
\vspace{-1cm}
\centering
\subfigure[$\;n=15$]{\label{F:QQa} \includegraphics[width=0.48\textwidth]{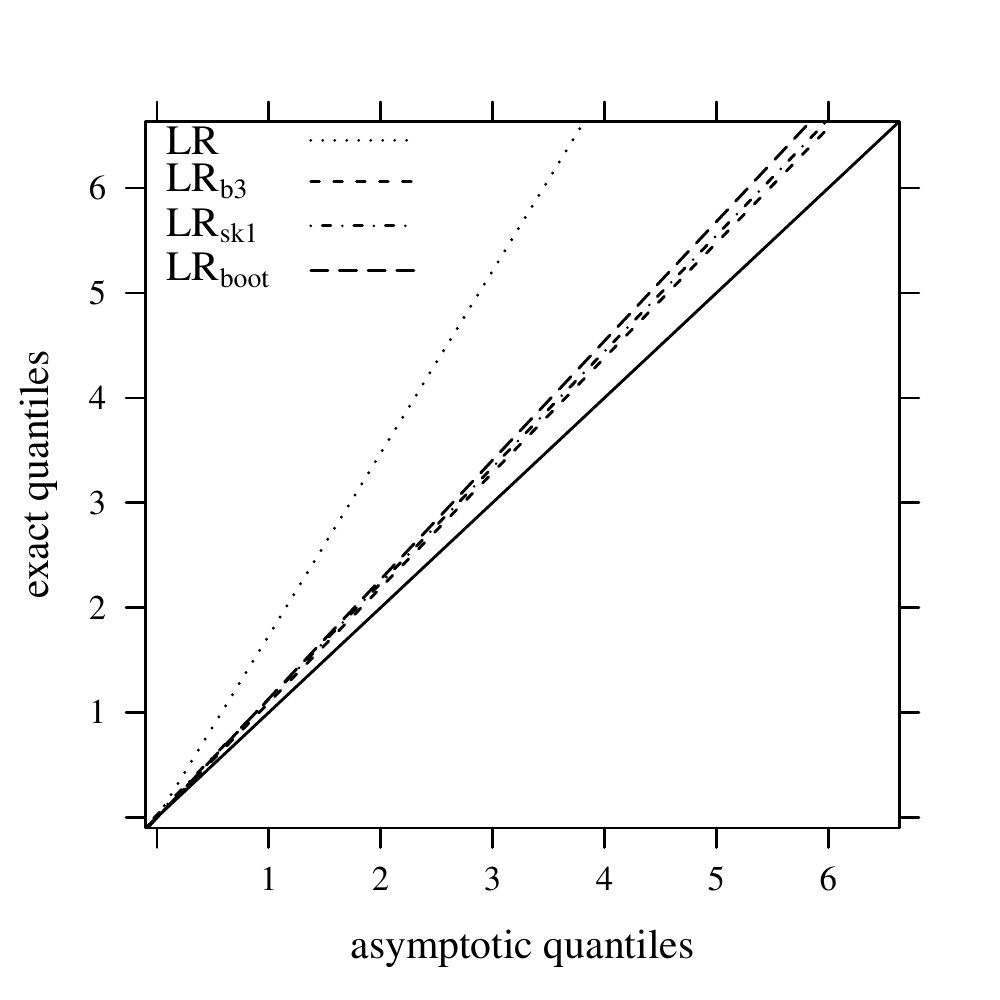}}
\subfigure[$\;n=20$] {\label{F:QQb} \includegraphics[width=0.48\textwidth] {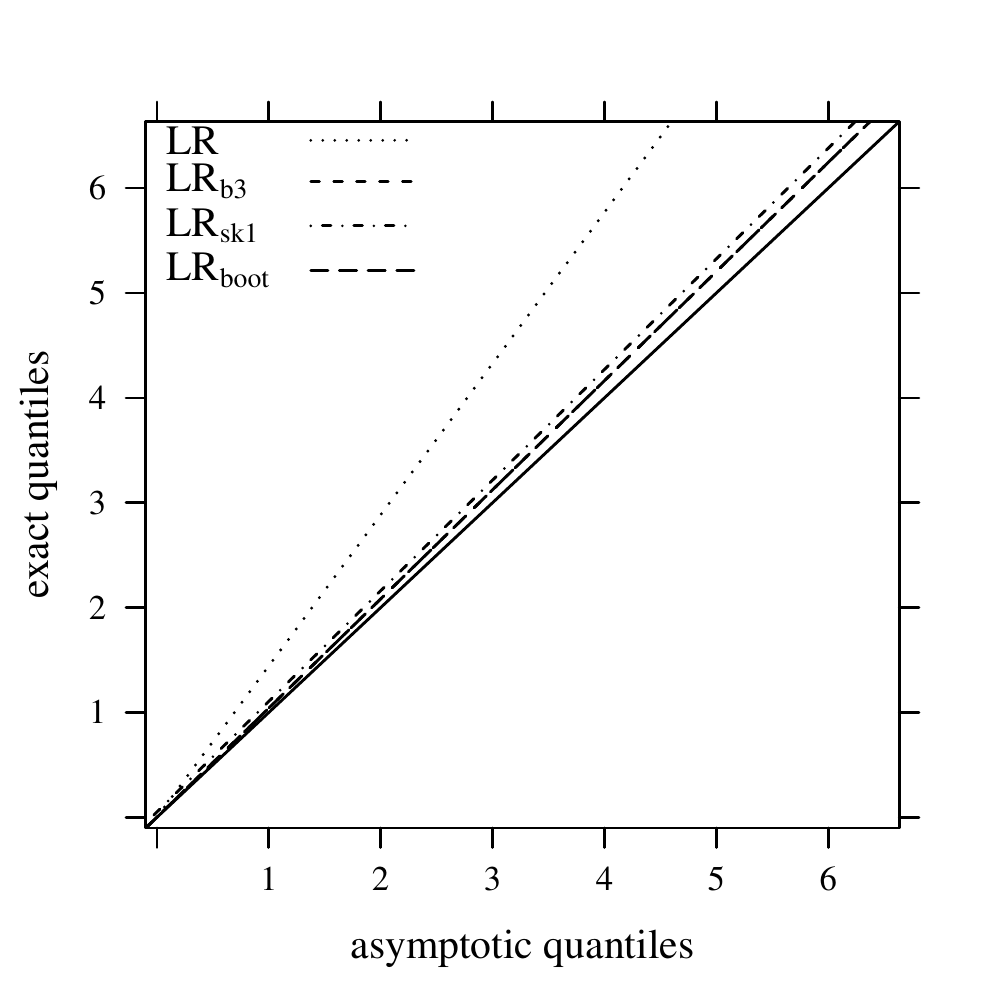}}
\subfigure[$\;n=30$] {\label{F:QQc} \includegraphics[width=0.48\textwidth] {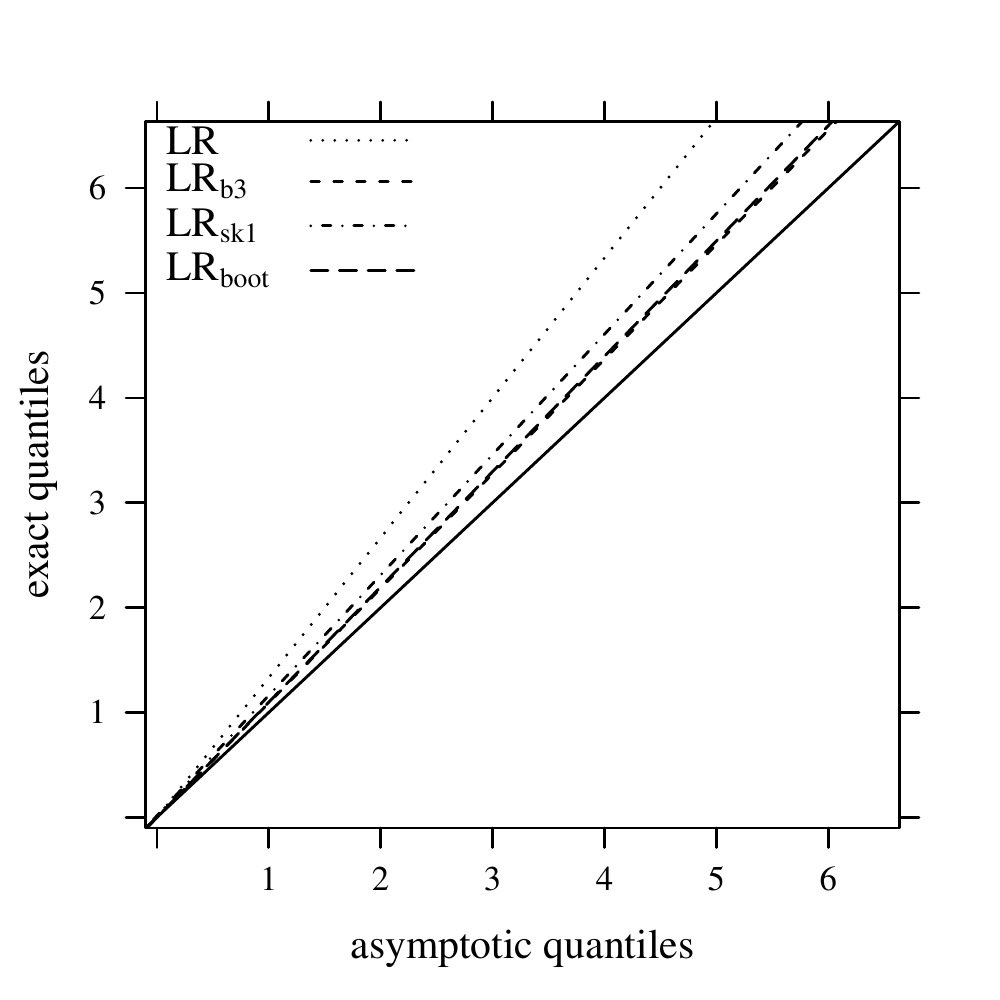}}
\subfigure[$\;n=40$] {\label{F:QQd} \includegraphics[width=0.48\textwidth] {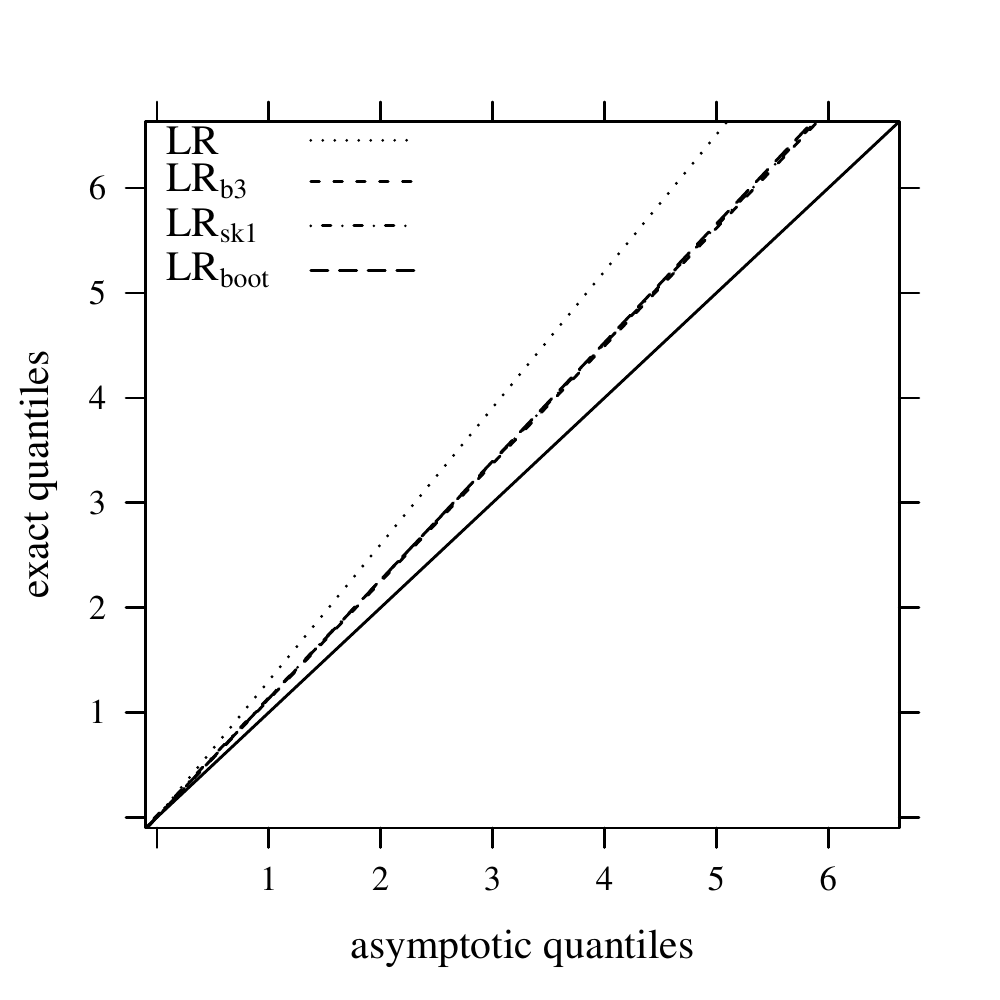}}
\caption{Quantile-quantile (QQ) plots for $\phi=10$ and $q=1$.}
\label{F:QQ}
\end{figure}

\begin{figure}[]
\vspace{-0.5cm}
\begin{center}
\subfigure[$\;n=15$]{\label{F:densitiesa}\includegraphics[width=0.9\textwidth]{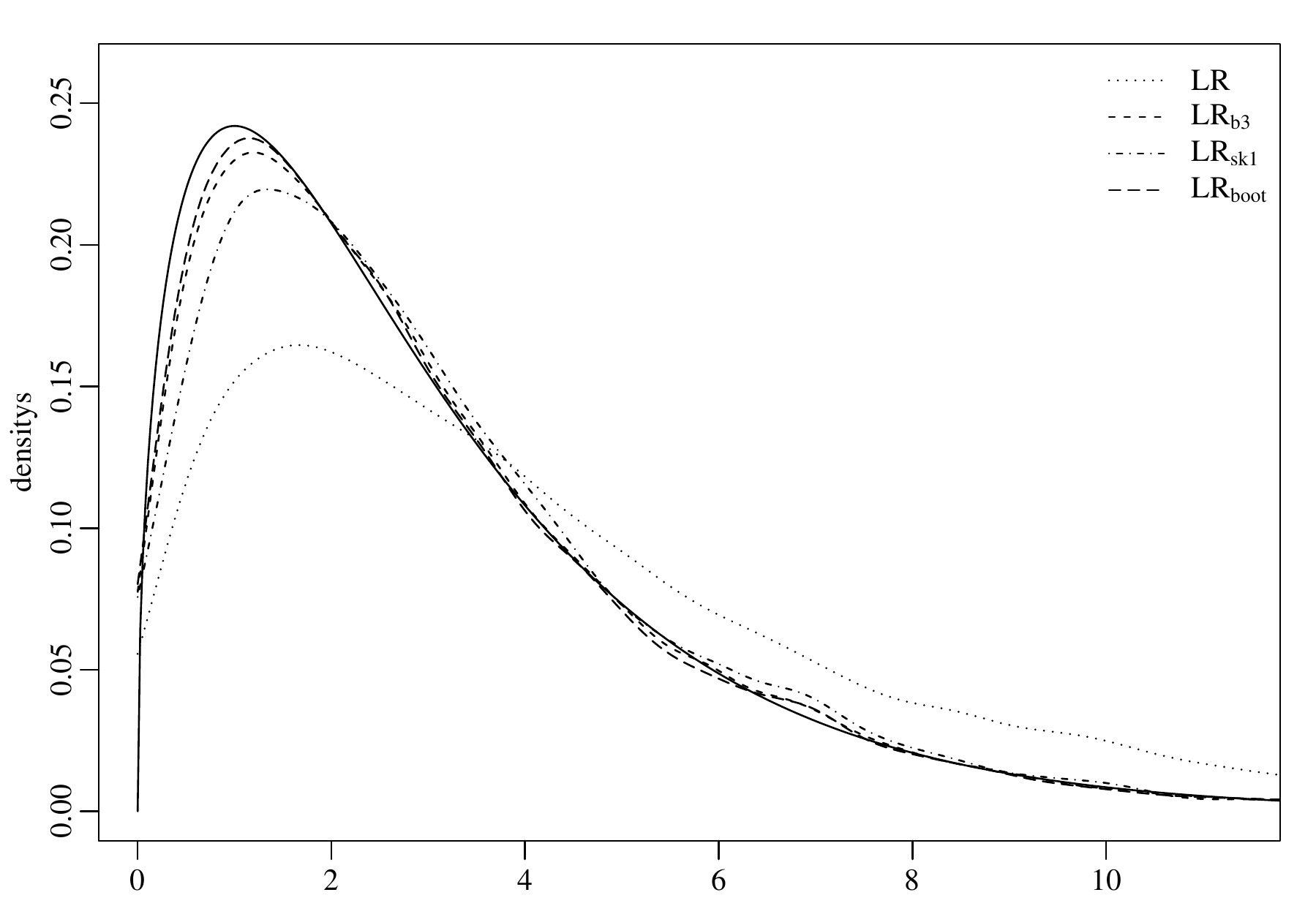}}
\subfigure[$\;n=20$]{\label{F:densitiesb}\includegraphics[width=0.9\textwidth]{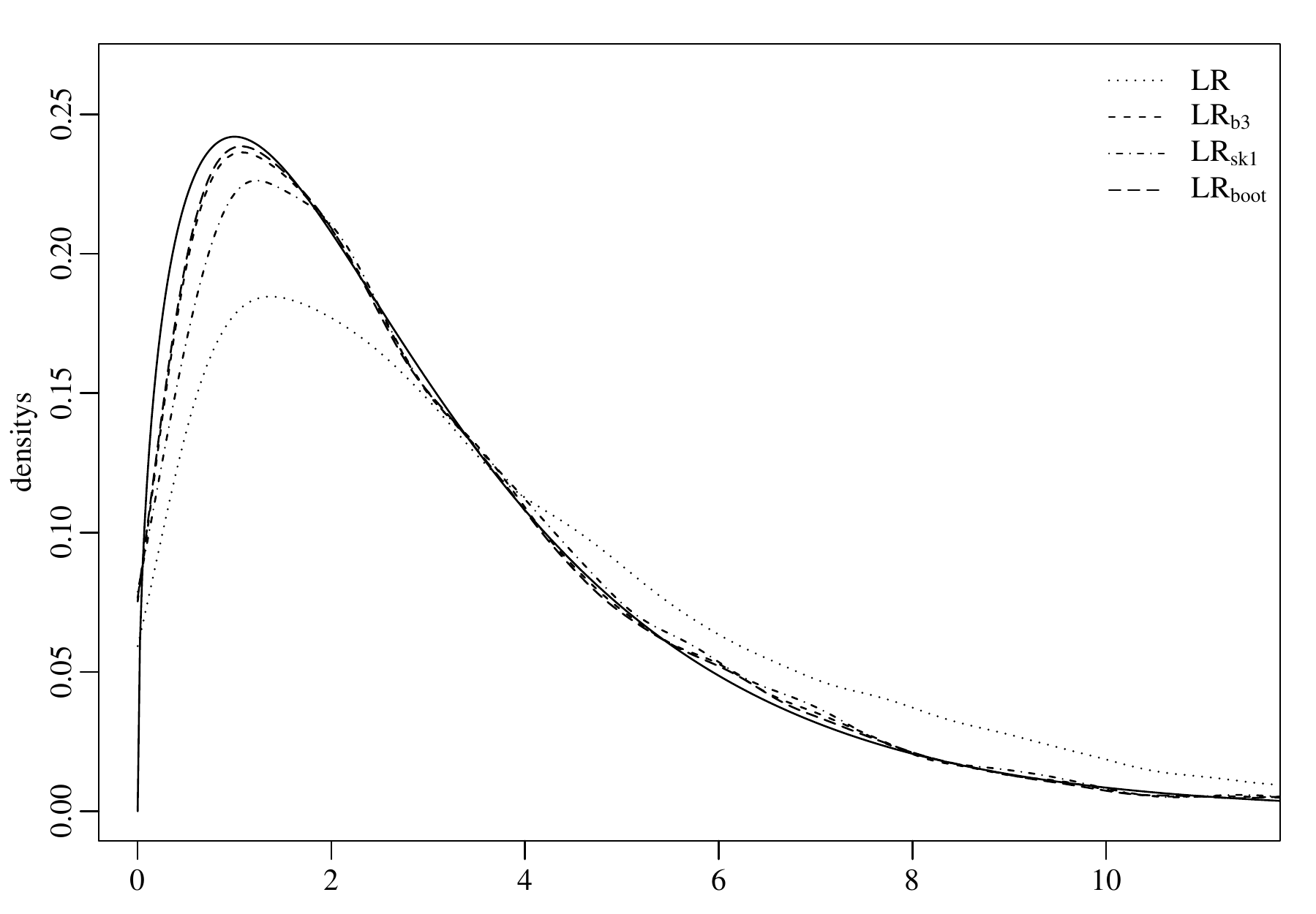}}			
\end{center}
\caption{$\chi^2_q$ density (solid line) and estimated null densities of four test statistics for $\phi=5$ and $q=3$.}
\label{F:densities}
\end{figure}

Figure~\ref{F:QQ} contains QQ plots (exact empirical quantiles versus asymptotic quantiles) for different sample sizes when $\phi=10$ and $q=1$. Figure \ref{F:densities} shows estimated null densities of some statistics for $\phi=5$ and $q=3$. These densities were estimated using the kernel method with Gaussian kernel function.\footnote{For details on nonparametric density estimation, see \cite{Silverman1986} and \cite{Venables2002}.} In both figures we consider the likelihood ratio test statistic, the best performer Bartlett corrected statistic ($LR_{b3}$), the bootstrap Bartlett corrected statistic and the best performer statistic modified using Skovgaard's approach ($LR_{sk1}$). The QQ plots in Figure~\ref{F:QQ} show that the corrected statistics null distributions are much more closer to the reference distribution than that of $LR$. The best agreement between exact and limiting null distributions takes place for $LR_{b3}$. The same conclusion can be drawn from the estimated null densities presented in Figure~\ref{F:densities}.

\begin{table}[!ht]

\caption{Nonnull rejection rates (\%); $n=20$ and $\alpha=5\%$.} 
\label{T:power}
{
\tablesize
\begin{center}
\begin{tabular}{ll|cccccccccc}
\hline
$\phi$ & \backslashbox[10pt][c]{Stat}{$\delta$} & $-2.5$ & $-2.0$ & $-1.5$ & $-1.0$ & $-0.5$ & $0.5$ & $1.0$ & $1.5$ & $2.0$ & $2.5$  \\
\hline
\multicolumn{12}{c}{$q=1$} \\
\hline
	 & $	LR_{b3}	$ & $	100	$ & $	100	$ & $	100	$ & $	99.5	$ & $	61.5$ & $	61.9$ & $	99.5	$ & $	100	$ & $	100		$ & $	100	$\\
100	 & $	LR_{sk1}$ & $	100	$ & $	100	$ & $	100	$ & $	99.4	$ & $	60.8$ & $	62.1$ & $	99.5	$ & $	100	$ & $	100		$ & $	100	$\\
	 & $LR_{boot}	$ & $	100	$ & $	100	$ & $	100	$ & $	99.5	$ & $	61.6$ & $	61.8$ & $	99.5	$ & $	100	$ & $	100		$ & $	100	$\\
\hline	
	 & $	LR_{b3}	$ & $	100		$ & $	99.8	$ & $	96.5	$ & $	71.1	$ & $	24.8	$ & $	25.2	$ & $	71.8	$ & $	96.3	$ & $	99.8	$ & $	100	$ \\
30	 & $	LR_{sk1}$ & $	100		$ & $	99.8	$ & $	96.4	$ & $	70.7	$ & $	24.7	$ & $	25.4	$ & $	72.3	$ & $	96.4	$ & $	99.8	$ & $	100	$ \\
	 & $LR_{boot}	$ & $	100		$ & $	99.8	$ & $	96.4	$ & $	71.1	$ & $	24.8	$ & $	25.1	$ & $	71.9	$ & $	96.1	$ & $	99.8	$ & $	100	$ \\
\hline
	 & $	LR_{b3}	$ & $	96.2	$ & $	85.3	$ & $	62.0	$ & $	33.3	$ & $	11.6	$ & $	12.5	$ & $	34.7	$ & $	63.5	$ & $	86.0	$ & $	96.2$\\
10	 & $	LR_{sk1}$ & $	96.0	$ & $	85.2	$ & $	61.7	$ & $	33.3	$ & $	11.5	$ & $	13.2	$ & $	36.0	$ & $	64.8	$ & $	86.8	$ & $	96.5$\\
	 & $LR_{boot}	$ & $	96.3	$ & $	85.6	$ & $	62.4	$ & $	33.7	$ & $	12.0	$ & $	12.7	$ & $	34.9	$ & $	63.6	$ & $	85.8	$ & $	96.2$\\
\hline
	 & $	LR_{b3}	$ & $	79.6	$ & $	61.2	$ & $	39.6	$ & $	20.4	$ & $	9.0		$ & $	9.6		$ & $	21.5	$ & $	41.4	$ & $	62.1	$ & $	81.2$\\
5	 & $	LR_{sk1}$ & $	79.8	$ & $	61.5	$ & $	39.8	$ & $	20.7	$ & $	9.2		$ & $	10.6	$ & $	23.2	$ & $	43.6	$ & $	64.4	$ & $	82.8$\\
	 & $LR_{boot}	$ & $	80.4	$ & $	62.1	$ & $	40.6	$ & $	21.4	$ & $	9.6		$ & $	9.8		$ & $	21.8	$ & $	41.4	$ & $	62.1	$ & $	80.9$\\
\hline
\multicolumn{12}{c}{$q=2$} \\
\hline	
	 & $	LR_{b3}	$ & $	100$ & $	100$ & $	100$ & $	100$ & $	79.6$ & $	80.4$ & $	100$ & $	100$ & $	100$ & $100$\\
100	 & $	LR_{sk1}$ & $	100$ & $	100$ & $	100$ & $	100$ & $	79.2$ & $	80.2$ & $	100$ & $	100$ & $	100$ & $100$\\
	 & $LR_{boot}	$ & $	100$ & $	100$ & $	100$ & $	100$ & $	79.2$ & $	80.3$ & $	100$ & $	100$ & $	100$ & $100$\\
\hline					
	 & $	LR_{b3}	$ & $	100		$ & $	100		$ & $	99.7	$ & $	87.7	$ & $	32.7	$ & $	31.6	$ & $	88.1	$ & $	99.7	$ & $	100		$ & $	100	$ \\
30	 & $	LR_{sk1}$ & $	100		$ & $	99.9	$ & $	99.7	$ & $	88.0	$ & $	32.9	$ & $	31.4	$ & $	88.1	$ & $	99.7	$ & $	100		$ & $	100	$ \\
	 & $LR_{boot}	$ & $	100		$ & $	99.9	$ & $	99.7	$ & $	87.7	$ & $	32.8	$ & $	31.6	$ & $	88.0	$ & $	99.6	$ & $	100		$ & $	100	$ \\
\hline
	 & $	LR_{b3}	$ & $	99.7	$ & $	97.6	$ & $	82.2	$ & $	47.2	$ & $	15.3	$ & $	15.6	$ & $	47.5	$ & $	82.5	$ & $	97.4	$ & $	99.7$\\
10	 & $	LR_{sk1}$ & $	99.8	$ & $	97.8	$ & $	82.9	$ & $	47.8	$ & $	15.5	$ & $	15.8	$ & $	48.0	$ & $	82.9	$ & $	97.6	$ & $	99.7$\\
	 & $LR_{boot}	$ & $	99.7	$ & $	97.6	$ & $	82.3	$ & $	47.2	$ & $	15.3	$ & $	15.5	$ & $	47.7	$ & $	82.2	$ & $	97.3	$ & $	99.7$\\
\hline
	 & $	LR_{b3}	$ & $	95.7	$ & $	84.1	$ & $	60.6	$ & $	25.8	$ & $	11.6	$ & $	9.5		$ & $	26.6	$ & $	53.8	$ & $	79.3	$ & $	93.7$\\
5	 & $	LR_{sk1}$ & $	96.0	$ & $	84.8	$ & $	61.5	$ & $	25.6	$ & $	11.8	$ & $	10.4	$ & $	27.9	$ & $	56.1	$ & $	81.0	$ & $	94.7$\\
	 & $LR_{boot}	$ & $	95.8	$ & $	84.5	$ & $	61.4	$ & $	26.2	$ & $	11.9	$ & $	9.7		$ & $	26.2	$ & $	53.5	$ & $	78.8	$ & $	93.5$\\
\hline
\multicolumn{12}{c}{$q=3$} \\
\hline	
	 & $	LR_{b3}	$ & $	100	$ & $	100	$ & $	100	$ & $	100	$ & $	92.3$ & $	91.3	$ & $	100	$ & $	100	$ & $	100	$ & $	100 $\\
100	 & $	LR_{sk1}$ & $	100	$ & $	100	$ & $	100	$ & $	100	$ & $	92.2$ & $	91.3	$ & $	100	$ & $	100	$ & $	100	$ & $	100	$\\
	 & $LR_{boot}	$ & $	100	$ & $	100	$ & $	100	$ & $	100	$ & $	92.2$ & $	91.4	$ & $	100	$ & $	100	$ & $	100	$ & $	100	$\\
\hline	
	 & $	LR_{b3}	$ & $	100		$ & $	100		$ & $	100		$ & $	96.2	$ & $	41.9	$ & $	41.6	$ & $	95.1	$ & $	99.9	$ & $	100		$ & $	100	$\\
30	 & $	LR_{sk1}$ & $	100		$ & $	100		$ & $	100		$ & $	96.0	$ & $	41.8	$ & $	41.5	$ & $	94.9	$ & $	100		$ & $	100		$ & $	100	$\\
	 & $LR_{boot}	$ & $	99.9	$ & $	100		$ & $	100		$ & $	96.2	$ & $	42.2	$ & $	41.5	$ & $	95.0	$ & $	100		$ & $	100		$ & $	100	$\\
\hline
	 & $	LR_{b3}	$ & $	100		$ & $	99.2	$ & $	91.5	$ & $	58.6	$ & $	17.8	$ & $	17.0	$ & $	56.9	$ & $	89.7	$ & $	98.5	$ & $	99.9$\\
10	 & $	LR_{sk1}$ & $	100		$ & $	99.2	$ & $	91.4	$ & $	58.3	$ & $	17.8	$ & $	17.3	$ & $	57.2	$ & $	89.7	$ & $	98.7	$ & $	99.9$\\
	 & $LR_{boot}	$ & $	100		$ & $	99.2	$ & $	91.4	$ & $	58.3	$ & $	17.7	$ & $	17.2	$ & $	56.9	$ & $	89.6	$ & $	98.5	$ & $	99.9$\\
\hline
	 & $	LR_{b3}	$ & $	98.0	$ & $	90.4	$ & $	69.4	$ & $	35.8	$ & $	11.3	$ & $	11.8	$ & $	35.6	$ & $	68.0	$ & $	88.7	$ & $	97.4$\\
5	 & $	LR_{sk1}$ & $	97.9	$ & $	90.2	$ & $	69.3	$ & $	35.7	$ & $	11.3	$ & $	12.3	$ & $	36.1	$ & $	68.5	$ & $	89.6	$ & $	97.7$\\
	 & $LR_{boot}	$ & $	97.9	$ & $	90.4	$ & $	69.6	$ & $	35.8	$ & $	11.2	$ & $	11.7	$ & $	35.3	$ & $	67.7	$ & $	88.4	$ & $	97.2$\\
\hline
\end{tabular}
\end{center}}
\end{table}

We have also used Monte Carlo simulation to estimate the tests nonnull rejection rates, i.e., their powers. Table~\ref{T:power} presents such rates when data generation was carried out using $\beta_2=\delta$ ($q=1$), $\beta_2=\beta_3=\delta$ ($q=2$) and $\beta_2=\beta_3=\beta_4=\delta$ ($q=3$), for different values of $\delta$. We only considered the corrected tests $LR_{b3}$, $LR_{boot}$ and $LR_{sk1}$. The likelihood ratio test is not included in the power comparison because it is considerably oversized.
Table~\ref{T:power} contains the estimated powers of the three tests for different values of $\delta$. As expected, the tests become more powerful as $\delta$ moves away from zero. We also note that the test based on $LR_{sk1}$ is slightly more powerful than the other two tests, especially when $\delta>0$. When $\delta<0$, $LR_{b3}$ is the most powerful test in some scenarios, e.g., when $\phi=5$ and $q=3$, as well as when $\phi=10$ and $q=1$. When $\delta=-0.5$, $q=1$, $\phi=10$ and $5$, $LR_{boot}$ outperforms the competition.

\section{An application}\label{S:application}

This section contains an application of the corrected likelihood ratio tests using data from a random sample of 38 households in a large U.S.\ city; the source of the data is \citeauthor{Griffiths1993}(1993, Tabela~15.4). \cite{Ferrari2004} fitted a beta regression model to these data. The response $(y)$ is the proportion of income spent on food and the covariates are income $(x_2)$ and the number of persons in the household $(x_3)$. We also consider as candidate covariates the interaction between income and number of persons $(x_4=x_2 \times x_3)$, income squared $(x_5=x_2^2)$ and the square of the number of persons in the household $(x_6=x_3^2)$. The beta regression model we fit is  
\begin{equation}\label{E:model_aplic_1}
{\rm logit}(\mu_i) = \beta_1 +\beta_2 x_{2i} +\beta_3 x_{3i} +\beta_4 x_{4i} +\beta_5 x_{5i} +\beta_6 x_{6i},
\end{equation}
where $i=1,\ldots,38$.

At the outset, we wish to make inference on the significance of the interaction variable ($x_4$), i.e., we wish to test $\mathcal{H}_0 : \beta_4=0$ against a two-sided alternative. The likelihood ratio test statistic ($LR$) equals 3.859 ($p$-value: 0.049) and the corrected test statistics are $LR_{b3}=3.208$ ($p$-value: 0.073) and $LR_{boot}=3.192$ ($p$-value: 0.074). These results show that inference is reversed when based on the corrected statistics. The likelihood ratio test rejects the null hypothesis at the 5\% nominal level whereas the two corrected tests yield a different conclusion at the same nominal level. 

We then remove the interaction variable ($x_4$) from the model and estimate the following reduced model: 
\begin{equation*}
{\rm logit}(\mu_i) = \beta_1 +\beta_2 x_{2i} +\beta_3 x_{3i} +\beta_5 x_{5i} +\beta_6 x_{6i}.
\end{equation*}
The point estimates are (standard errors in parentheses): $\hat{\beta}_1 = 0.4861$ (0.5946), $\hat{\beta}_2 = -0.0495$ (0.0218), $\hat{\beta}_3 = 0.0172$ (0.1563), $\hat{\beta}_5 = 0.0003$ (0.0002), $\hat{\beta}_6 = 0.0129$ (0.0198) and $\hat{\phi}=39.296$ (8.925). We now wish to test $\mathcal{H}_0 : \beta_5=\beta_6=0$. The statistics are $LR=3.791$ ($p$-value: 0.150), $LR_{b3}=3.296$ ($p$-value: 0.192) and $LR_{boot}=3.210$ ($p$-value: 0.201). The null hypothesis is not rejected by the three tests at the usual nominal levels. 

We thus arrive at the following reduced model:
\begin{equation*}
{\rm logit}(\mu_i) = \beta_1 +\beta_2 x_{2i} +\beta_3 x_{3i}.
\end{equation*}
The point estimates (standard errors in parentheses) are $\hat{\beta}_1 = -0.6225$ (0.224), $\hat{\beta}_2 = -0.0123$ (0.003), $\hat{\beta}_3 = 0.1185$ (0.035) and $\hat{\phi}=35.61$ (8.080).

We now return to the Model (\ref{E:model_aplic_1}) and test the joint exclusion of the three regressors, i.e., we test $\mathcal{H}_0 : \beta_4=\beta_5=\beta_6=0$. For this test we obtain $LR=7.6501$ ($p$-value: 0.054), $LR_{b3}=6.554$ ($p$-value: 0.088) and $LR_{boot}=6.068$ ($p$-value: 0.108). The $p$-value of the unmodified test is very close to 5\% whereas the $p$-values of the corrected tests indicate that the null rejection is not to be rejected at the 5\% nominal level. It is noteworthy that the null hypothesis is not rejected by the bootstrap Bartlett corrected test at the 10\% nominal level.

\section{Conclusions}

The class of beta regression models is commonly used when the interest lies in modeling the behavior of variables that assume values in the standard unit interval, such as rates and proportions. Testing inference is typically performed via the likelihood ratio test which is performed using asymptotic critical values, i.e., critical values obtained from the test statistic limiting null distribution. When the sample size is small, however, the approximation tends to be poor and size distortions take place. It is thus important to develop testing inference strategies that are more accurate when the sample size is not large. \cite{Pinheiro2011} derived two modified likelihood ratio test statistics for testing restrictions on beta regressions that typically yield more reliable inferences. They considered a very general class of models, which allows for nonlinearities and varying dispersion. In this paper, we derived three Bartlett corrected likelihood ratio test statistics for fixed dispersion beta regressions. The derivation is considerably more cumbersome than that of \cite{Pinheiro2011}, especially because $\beta$ and $\phi$ are not orthogonal. A clear advantage of our approach is that it delivers tests with higher order of accuracy. That is, the size distortions of our tests vanish faster than those of the unmodified likelihood ratio test and also than those of the modified tests proposed by \cite{Pinheiro2011}. We also considered a different approach in which bootstrap data resampling is used to estimate the Bartlett correction factor. We reported results of Monte Carlo simulations that show that the likelihood ratio test tends to be quite liberal (oversized) in small samples. The numerical evidence also shows that the corrected tests deliver much more accurate testing inference. In particular, one of the analytically derived Bartlett corrected tests ($LR_{b3}$) and the bootstrap Bartlett corrected test display superior finite sample behavior. We strongly encourage practitioners to base inference on such tests when performing beta regression analyses.

\section*{Acknowledgements}

We gratefully acknowledge partial financial support from Coordenação de Aperfeiçoamento de Pessoal de Nível Superior (CAPES) and Conselho Nacional de Desenvolvimento Científico e Tecnológico (CNPq), Brazil.

\appendix

\section{Cumulants for the Bartlett correction factor} \label{A:cumulants}

In this appendix we present the derivatives of the log-likelihood function in (\ref{E:loglik}) up to the fourth order with respect to the unknown parameters and obtain their moments. Cumulants up to the third order can be found in \cite{Ospina2006}.

At the outset, we define the following quantities:
\begin{align*}
& \omega_i=\psi'(\mu_i \phi) + \psi'((1-\mu_i) \phi),\\
& m_i= \psi''(\mu_i \phi) - \psi''((1-\mu_i) \phi),\\
& a_i= 3\left(\frac{\partial}{\partial\mu_i} \frac{d\mu_i}{d\eta_i}\right) \left(\frac{d\mu_i}{d\eta_i}\right)^2,\\
& b_i= \frac{d\mu_i}{d\eta_i}\left[\left(\frac{\partial^2}{\partial\mu_i^2} \frac{d\mu_i}{d\eta_i}\right)\frac{d\mu_i}{d\eta_i} + \left(\frac{\partial}{\partial\mu_i}\frac{d\mu_i}{d\eta_i}\right)^2\right],\\
& c_i = \phi\left[\mu_i\omega_i - \psi'((1-\mu_i)\phi)\right]= \phi\frac{\partial\mu_i^{*}}{\partial\phi},\\
& d_i = (1-\mu_i)^2 \psi'((1-\mu_i)\phi) + \mu_i^2 \psi'(\mu_i\phi) - \psi'(\phi),\\
& s_i = (1-\mu_i)^3 \psi''((1-\mu_i)\phi) + \mu_i^3 \psi''(\mu_i\phi)- \psi''(\phi),\\
& u_i = -\phi\left[2\omega_i + \phi\frac{\partial\omega_i}{\partial\phi}\right],\\
& r_i = \left[2\frac{\partial\mu_i^*}{\partial\phi} + \phi\frac{\partial^2\mu_i^*}{\partial\phi^2} \right]\frac{d\mu_i}{d\eta_i}.
\end{align*}
Closed-form expressions for $\partial\omega_i/\partial\phi$, $\partial\mu_i^*/\partial\phi$ and $\partial^2\mu_i^*/\partial\phi^2$ are given below. Additionally, we have 
\begin{align*}
& \frac{d \mu_i}{d\eta_i} =  \frac{1}{g'(\mu_i)},\\
& \frac{\partial}{\partial\mu_i}\frac{d\mu_i}{d\eta_i} =  \frac{-g''(\mu_i)}{(g'(\mu_i))^2},\\
& \frac{\partial}{\partial\mu_i}\left(\frac{d\mu_i}{d\eta_i}\right)^2 =  \frac{-2g''(\mu_i)}{(g'(\mu_i))^3},\\
& \frac{\partial}{\partial\mu_i}\left(\frac{d\mu_i}{d\eta_i}\right)^3 =  \frac{-3g''(\mu_i)}{(g'(\mu_i))^4},\\
& \frac{\partial^2}{\partial\mu_i^2}\left(\frac{d\mu_i}{d\eta_i}\right) =  \frac{-g'''(\mu_i)g'(\mu_i) + 2(g''(\mu_i))^2}{(g'(\mu_i))^3}.
\end{align*}
In particular, if the link function is logit, i.e., 
\begin{equation*}
g(\mu_i) = {\rm logit}(\mu_i) = \log\frac{\mu_i}{(1-\mu_i)},
\end{equation*}
as in our numerical evaluation, it follows that 
\begin{align*}
& g'(\mu_i) = \frac{1}{\mu_i(1-\mu_i)},\\
& g''(\mu_i) = \frac{2\mu_i - 1}{\mu_i^2(1-\mu_i)^2},\\
& g'''(\mu_i) = \frac{2(1 - 4\mu_i + 6\mu_i^2 - 3\mu_i^3)}{\mu_i^3(1-\mu_i)^4}.
\end{align*}

We also obtain the following derivatives:
\begin{align*}
& \frac{\partial\mu^*_i}{\partial\mu_i} = \phi\psi'(\mu_i\phi) + \phi\psi'((1-\mu_i)\phi) = \phi\omega_i,\\
& \frac{\partial\mu^*_i}{\partial\phi} = \mu_i\psi'(\mu_i\phi) - (1-\mu_i)\psi'((1-\mu_i)\phi) = \frac{c_i}{\phi},\\
& \frac{\partial\omega_i}{\partial\mu_i} = \phi\psi''(\mu_i\phi) - \phi\psi''((1-\mu_i)\phi) = \phi m_i,\\
& \frac{\partial\omega_i}{\partial\phi} = \mu_i\psi''(\mu_i\phi) + (1-\mu_i)\psi''((1-\mu_i)\phi),\\
& \frac{\partial m_i}{\partial\mu_i} = \phi\psi'''(\mu_i\phi) + \phi\psi'''((1-\mu_i)\phi),\\
& \frac{\partial m_i}{\partial\phi} = \mu_i\psi'''(\mu_i\phi) - (1-\mu_i)\psi'''((1-\mu_i)\phi),\\
& \frac{\partial a_i}{\partial\mu_i} = 3\left(\frac{d\mu_i}{d\eta_i} \right)\left[\left(\frac{\partial^2}{\partial\mu_i^2}\frac{d\mu_i}{d\eta_i} \right)\left(\frac{d\mu_i}{d\eta_i} \right) + 2\left(\frac{\partial}{\partial\mu_i}\frac{d\mu_i}{d\eta_i} \right)^2 \right],\\
& \frac{\partial b_i}{\partial\mu_i} = \left(\frac{\partial}{\partial\mu_i}\frac{d\mu_i}{d\eta_i} \right)^3 + \left(\frac{d\mu_i}{d\eta_i} \right)\left[ \left(\frac{\partial^3}{\partial\mu_i^3}\frac{d\mu_i}{d\eta_i} \right)\left(\frac{d\mu_i}{d\eta_i} \right) + 4\left(\frac{\partial^2}{\partial\mu_i^2}\frac{d\mu_i}{d\eta_i} \right)\left(\frac{\partial}{\partial\mu_i}\frac{d\mu_i}{d\eta_i} \right)\right],\\
& \frac{\partial^2\mu^*_i}{\partial\phi^2} = \mu_i^2\psi''(\mu_i\phi) - (1-\mu_i)^2\psi''((1-\mu_i)\phi),\\
& \frac{\partial^2\omega_i}{\partial\phi^2} = \mu_i^2\psi'''(\mu_i\phi) + (1-\mu_i)^2\psi'''((1-\mu_i)\phi),\\
& \frac{\partial^3\mu^*_i}{\partial\phi^3} = \mu_i^3\psi'''(\mu_i\phi) - (1-\mu_i)^3\psi'''((1-\mu_i)\phi),\\
& \frac{\partial c_i}{\partial\mu_i} = \phi \left(\omega_i + \phi\frac{\partial\omega_i}{\partial\phi} \right),\\
& \frac{\partial c_i}{\partial\phi_i} = \frac{\partial\mu_i^*}{\partial\phi} + \phi\frac{\partial^2\mu_i^*}{\partial\phi^2},\\
& \frac{\partial s_i}{\partial\mu_i} = 3\frac{\partial^2\mu_i^*}{\partial\phi^2} + \phi\frac{\partial^3\mu_i^*}{\partial\phi^3},\\
& \frac{\partial s_i}{\partial\phi_i} = \mu_i^4\psi'''(\mu_i\psi) + (1-\mu_i)^4\psi'''((1-\mu_i)\phi) - \psi'''(\phi),\\
& \frac{\partial u_i}{\partial\mu_i} = -\phi^2 \left(3m_i + \phi\frac{\partial m_i}{\partial\phi} \right),\\
& \frac{\partial u_i}{\partial\phi_i} = -2\omega_i - \phi\left(4\frac{\partial\omega_i}{\partial\phi} + \frac{\partial^2\omega_i}{\partial\phi^2} \right),\\
& \frac{\partial r_i}{\partial\mu_i} = \left(2\frac{\partial\mu_i^*}{\partial\phi} + \phi\frac{\partial^2\mu_i^*}{\partial\phi^2} \right)\left(\frac{\partial}{\partial\mu_i}\frac{d\mu_i}{d\eta_i} \right) +\left(2\omega_i + 4\phi\frac{\partial\omega_i}{\partial\phi} + \phi^2\frac{\partial^2\omega_i}{\partial\phi^2} \right)\left(\frac{d\mu_i}{d\eta_i} \right),\\
& \frac{\partial r_i}{\partial\phi_i} = \left(3\frac{\partial^2\mu_i^*}{\partial\phi^2} + \phi\frac{\partial^3\mu_i^*}{\partial\phi^3} \right)\frac{d\mu_i}{d\eta_i} = \frac{\partial s_i}{\partial\mu_i}\frac{d\mu_i}{d\eta_i},\\
& \frac{\partial}{\partial\mu_i}\frac{\partial\omega_i}{\partial\phi} = m_i + \phi\frac{\partial m_i}{\partial\phi}.
\end{align*}

Using the above expressions, the second, third and forth order derivatives of the log-likelihood function are
\begin{align*}
& U_{rs} = \sum_{i=1}^n{\left\{-\phi^2\omega_i\left(\frac{d\mu_i}{d\eta_i} \right)^2 + \phi\left[y_i^* - \mu_i^* \right]\left(\frac{\partial}{\partial\mu_i}\frac{d\mu_i}{d\eta_i} \right)\frac{d\mu_i}{d\eta_i} \right\}}x_{ir}x_{is},\\
& U_{r\phi} =- \sum_{i=1}^n{\left[c_t - (y_i^* - \mu_i^*) \right]}\frac{d\mu_i}{d\eta_i}x_{ir},\\
& U_{\phi\phi} =- \sum_{i=1}^n{d_i},\\
& U_{rst} =-\phi \sum_{i=1}^n{\left\{\phi^2m_i\left(\frac{d\mu_i}{d\eta_i} \right)^3 + \phi\omega_ia_i - \left[y_i^* - \mu_i^* \right]b_i \right\}}x_{ir}x_{is}x_{it},\\
& U_{rs\phi} =\sum_{i=1}^n{\left\{\mu_i\left(\frac{d\mu_i}{d\eta_i} \right) + \left(y_i^* - \mu_i^* \right)\left(\frac{\partial}{\partial\mu_i}\frac{d\mu_i}{d\eta_i} \right) - c_i\left(\frac{\partial}{\partial\mu_i}\frac{d\mu_i}{d\eta_i} \right)\right\}\frac{d\mu_i}{d\eta_i}}x_{ir}x_{is},\\
& U_{r\phi\phi} =- \sum_{i=1}^n{r_t}x_{ir},\\
& U_{\phi\phi\phi} =- \sum_{i=1}^n{s_i},\\
& U_{rstu} = -\phi \sum_{i=1}^n \left\{
\phi^2 
\left[ m_i \frac{\partial}{\partial \mu_i}\left(\frac{d\mu_i}{d\eta_i}\right)^3 + \frac{\partial m_i}{\partial\mu_i} \left(\frac{d\mu_i}{d\eta_i}\right)^3 \right] \right.\\
& \quad \quad \left.+ \, \phi
\left[ \left( \frac{\partial a_i}{\partial \mu_i} +  b_i \right) \omega_i + \frac{\partial \omega_i}{\partial \mu_i} a_i \right] 
- \left(y_i^*-\mu_i^* \right)\frac{\partial b_i}{\partial \mu_i} 
\right\}\frac{d\mu_i}{d\eta_i} x_{ir}x_{is}x_{it}x_{iu},\\
& U_{rst\phi}  = \sum_{i=1}^n \left\{-\phi \left[\phi \left(3 m_t + \phi^2 \frac{\partial m_i}{\partial \phi}\right)\left(\frac{d\mu_i}{d\eta_i}\right)^3 +
a_i\left(2 \omega_i + \phi \frac{\partial \omega_i}{\partial \phi} \right) +
b_i\frac{\partial \mu_i^*}{\partial \phi} \right] \right. \\
& \quad \quad \left. + \,b_i\left(y_i^*-\mu_i^* \right) \frac{}{}\right\} x_{ir}x_{is}x_{it},\\
& U_{rs\phi\phi} = - \sum_{i=1}^n \frac{\partial r_i}{\partial \mu_i} \frac{d\mu_i}{d\eta_i}x_{ir}x_{is},\\
& U_{r\phi\phi\phi} = - \sum_{i=1}^n \frac{\partial s_i}{\partial \mu_i} \frac{d\mu_i}{d\eta_i}x_{ir},\\
& U_{\phi\phi\phi\phi} = - \sum_{i=1}^n \frac{\partial s_i}{\partial \phi}.
\end{align*}


By taking the expected values of the above derivatives, we obtain the following cumulants: 
\begin{align*}
& \kappa_{rs} = -\phi^2 \sum_{i=1}^n{\omega_i\left(\frac{d\mu_i}{d\eta_i} \right)^2 x_{ir}x_{is}},\\
& \kappa_{r\phi} = -\sum_{i=1}^n{c_i\frac{d\mu_i}{d\eta_i}x_{ir}},\\
& \kappa_{\phi\phi} = -\sum_{i=1}^n{d_i},\\
& \kappa_{rst} = -\phi^2 \sum_{i=1}^n{\left[\phi m_i\left(\frac{d\mu_i}{d\eta_i} \right)^3 + \omega_i a_i\right] x_{ir}x_{is}x_{it}},\\
& \kappa_{rs\phi} = \sum_{i=1}^n{\left[u_i\left(\frac{d\mu_i}{d\eta_i} \right)- c_i\left(\frac{\partial}{\partial\mu_i}\frac{d\mu_i}{d\eta_i}\right)\right] x_{ir}x_{is}},\\
& \kappa_{r\phi\phi} = -\sum_{i=1}^n{r_ix_{ir}},\\
& \kappa_{\phi\phi\phi} = -\sum_{i=1}^n{s_i},\\
& \kappa_{rstu} = -\phi^2 \sum_{i=1}^n\left\{\phi\left[m_i\frac{\partial}{\partial\mu_i}\left(\frac{d\mu_i}{d\eta_i} \right)^3 + \frac{\partial m_i}{\partial\mu_i}\left(\frac{d\mu_i}{d\eta_i} \right)^3 + m_ia_i \right] \right.\\
& \quad \quad \left.+ \, \omega_i\left(\frac{\partial a_i}{\partial\mu_i} +b_i \right) \right\}\frac{d\mu_i}{d\eta_i}x_{ir}x_{is}x_{it}x_{iu},\\
& \kappa_{rst\phi} = -\phi \sum_{i=1}^n{\left[\phi\left(3m_i + \phi\frac{\partial m_i}{\partial\phi}\right)\left(\frac{d\mu_i}{d\eta_i} \right)^3 + a_i\left(2\omega_i + \phi\frac{\partial \omega_i}{\partial\phi}\right) + b_i\frac{c_i}{\phi} \right]x_{ir}x_{is}x_{it}},\\
& \kappa_{rs\phi\phi} = -\sum_{i=1}^n{\frac{\partial r_i}{\partial\mu_i}\frac{d\mu_i}{d\eta_i}x_{ir}x_{is}},\\
& \kappa_{r\phi\phi\phi} = -\sum_{i=1}^n{\frac{\partial s_i}{\partial\mu_i}\frac{d\mu_i}{d\eta_i}x_{ir}},\\
& \kappa_{\phi\phi\phi\phi} = -\sum_{i=1}^n{\frac{\partial s_i}{\partial\phi}}.\\
\end{align*} 

Let
\begin{equation*}
 z_i = \frac{\partial\mu_i^*}{\partial\phi} + \phi\frac{\partial^2\mu_i^*}{\partial\phi^2}. 
 \end{equation*}
Hence, 
\begin{equation*}
\frac{\partial z_i}{\partial\mu_i} = \omega_i + \phi\left(3\frac{\partial\omega_i}{\partial\phi} + \phi\frac{\partial^2\omega_i}{\partial\phi^2}\right) 
\end{equation*}
and
\begin{equation*}
\frac{\partial z_i}{\partial\phi_i} = 2\frac{\partial^2\mu_i^*}{\partial\phi^2} + \phi\frac{\partial^3\mu_i^*}{\partial\phi^3}.
\end{equation*}

Differentiating the cumulants, we obtain
\begin{align*}
& \kappa_{rs}^{(u)} = -\phi^2 \sum_{i=1}^n{\left[\phi m_i\left(\frac{\partial \mu_i}{\partial\eta_i}\right)^3 +\frac{2}{3}\omega_ia_i\right]x_{ir}x_{is}x_{iu}},\\
& \kappa_{rs}^{(\phi)} = \sum_{i=1}^n{\left[u_i\left(\frac{\partial \mu_i}{\partial\eta_i}\right)^2\right]x_{is}x_{is}},\\
& \kappa_{r\phi}^{(u)} = - \sum_{i=1}^n{\left\{\left[\phi \left(\omega_i + \phi\frac{\partial \omega_i}{\partial\phi} \right) \right]\frac{d\mu_i}{d\eta_i} + c_i\left(\frac{\partial}{\partial\mu_i}\frac{\partial\mu_i}{\partial\eta_i} \right) \right\}\frac{d\mu_i}{d\eta_i}x_{ir}x_{iu}},\\
& \kappa_{r\phi}^{(\phi)} = - \sum_{i=1}^n{z_i\frac{d\mu_i}{d\eta_i}x_{ir}},\\
& \kappa_{\phi\phi}^{(u)} = - \sum_{i=1}^n{r_ix_{iu}},\\
& \kappa_{\phi\phi}^{(\phi)} = - \sum_{i=1}^n{s_i},\\
& \kappa_{rst}^{(u)} = - \phi^2\sum_{i=1}^n{\left\{\phi\left[m_i\left(\frac{\partial}{\partial\mu_i}\left(\frac{d\mu_i}{d\eta_i} \right)^3 + a_i \right) + \left(\frac{d\mu_i}{d\eta_i} \right)^3 \frac{\partial m_i}{\partial\mu_i} \right] + \omega_i\frac{\partial a_i}{\partial\mu_i} \right\}\frac{d\mu_i}{d\eta_i}x_{ir}x_{is}x_{it}x_{iu}},\\
& \kappa_{rst}^{(\phi)} = - \phi\sum_{i=1}^n{\left[\left(\frac{d\mu_i}{d\eta_i} \right)^3 \left( 3\phi m_i + \phi^2\frac{\partial m_i}{\partial\phi}\right) + a_i\left(2\omega_i+ \phi\frac{\partial\omega_i}{\partial\phi} \right)\right]x_{ir}x_{is}x_{it}},\\
& \kappa_{rs\phi}^{(t)} =\sum_{i=1}^n \left\{\frac{\partial u_i}{\partial\mu_i}\left(\frac{d\mu_i}{d\eta_i} \right)^2 + \mu_i\frac{\partial}{\partial\mu_i}\left(\frac{d\mu_i}{d\eta_i} \right)^2 - \frac{\partial c_i}{\partial\mu_i}\left(\frac{\partial}{\partial\mu_i}\frac{d\mu_i}{d\eta_i} \right)\frac{d\mu_i}{d\eta_i} \right.\\
& \quad \quad - \left. \phi\frac{\partial\mu_i^*}{\partial\phi}\left[\left(\frac{\partial^2}{\partial\mu_i^2}\frac{d\mu_i}{d\eta_i} \right)\left(\frac{d\mu_i}{d\eta_i} \right) + \left(\frac{\partial}{\partial\mu_i}\frac{d\mu_i}{d\eta_i} \right)^2 \right] \right\}x_{ir}x_{is}x_{it},\\
& \kappa_{rs\phi}^{(\phi)} =\sum_{i=1}^n{\left[\left(\frac{d\mu_i}{d\eta_i} \right) \frac{\partial u_i}{\partial\phi} - \left(\frac{\partial}{\partial\mu_i}\frac{d\mu_i}{d\eta_i} \right) z_i \right]\frac{d\mu_i}{d\eta_i} x_{ir}x_{is}},\\
& \kappa_{r\phi\phi}^{(s)} =-\sum_{i=1}^n{\frac{\partial r_i}{\partial\mu_i}\frac{d\mu_i}{d\eta_i} x_{ir}x_{is}},\\
& \kappa_{r\phi\phi}^{(\phi)} =-\sum_{i=1}^n{\frac{\partial r_i}{\partial\phi} x_{ir}},\\
& \kappa_{\phi\phi\phi}^{(r)} =-\sum_{i=1}^n{\frac{\partial s_i}{\partial\mu_i}\frac{d\mu_i}{d\eta_i} x_{ir}},\\
& \kappa_{\phi\phi\phi}^{(\phi)} =-\sum_{i=1}^n{\frac{\partial s_i}{\partial\phi}}.
\end{align*} 

By taking second order derivatives of the second order cumulants, we obtain
\begin{align*}
& \kappa_{rs}^{(tu)} =-\phi^2\sum_{i=1}^n\left\{\phi\left[m_i \left(\frac{\partial}{\partial\mu_i}\left(\frac{d\mu_i}{d\eta_i}\right)^3 + \frac{2}{3} a_i \right) + \left(\frac{d\mu_i}{d\eta_i}\right)^3 \frac{\partial m_i}{\partial\mu_i} \right] \right.\\
&  \quad \quad \left. + \frac{2}{3}\omega_i \frac{\partial a_i}{\partial\mu_i} \right\}\frac{d\mu_i}{d\eta_i}x_{ir}x_{is}x_{it}x_{iu},\\
& \kappa_{rs}^{(tu)} =\sum_{i=1}^n{\left[\frac{\partial u_i}{\partial\mu_i}\frac{d\mu_i}{d\eta_i} + 2u_i\frac{\partial}{\partial\mu_i}\left(\frac{d\mu_i}{d\eta_i} \right)\right]\left(\frac{d\mu_i}{d\eta_i}\right)^2x_{ir}x_{is}x_{it}},\\
& \kappa_{rs}^{(\phi\phi)} =\sum_{i=1}^n{\frac{\partial\mu_i}{\partial\phi}\left(\frac{d\mu_i}{d\eta_i} \right)^2 x_{ir}x_{is}},\\
& \kappa_{r\phi}^{(st)} =-\phi\sum_{i=1}^n\left\{\frac{\partial\omega_i}{\partial\mu_i}\left(\frac{d\mu_i}{d\eta_i}\right)^2 + \omega_i\frac{\partial}{\partial\mu_i}\left(\frac{d\mu_i}{d\eta_i}\right)^2 \right.\\
& \quad \quad + \,\phi\left[\frac{\partial}{\partial\mu_i}\left(\frac{\partial\omega_i}{\partial\phi}\right)\left(\frac{d\mu_i}{d\eta_i}\right)^2 + \frac{\partial\omega_i}{\partial\phi}\frac{\partial}{\partial\mu_i}\left(\frac{d\mu_i}{d\eta_i}\right)^2\right] 
+ \left(\!\omega_i + \phi \frac{\partial\omega_i}{\partial\phi}\right)\! \left(\frac{\partial}{\partial\mu_i}\frac{d\mu_i}{d\eta_i}\right)\frac{d\mu_i}{d\eta_i} \\
& \quad \quad \left.+ \,\frac{\partial\mu_i^*}{\partial\phi}\left[\frac{\partial^2}{\partial\mu_i}\left(\frac{d\mu_i}{d\eta_i}\right)\left(\frac{d\mu_i}{d\eta_i}\right) + \left(\frac{\partial}{\partial\mu_i}\frac{d\mu_i}{d\eta_i}\right)^2\right] \right\}\frac{d\mu_i}{d\eta_i}x_{ir}x_{is}x_{it} ,\\
& \kappa_{r\phi}^{(\phi s)} =-\sum_{i=1}^n{\left[\frac{\partial z_i}{\partial\mu_i}\left(\frac{d\mu_i}{d\eta_i}\right) + z_i\left(\frac{\partial}{\partial\mu_i}\frac{d\mu_i}{d\eta_i}\right)\right]\frac{d\mu_i}{d\eta_i} x_{ir}x_{is}},\\
& \kappa_{r\phi}^{(\phi\phi)} =-\sum_{i=1}^n{\frac{\partial z_i}{\partial\phi}\frac{d\mu_i}{d\eta_i} x_{ir}x_{is}},\\
& \kappa_{\phi\phi}^{(rs)} =-\sum_{i=1}^n{\frac{\partial r_i}{\partial\mu_i}\frac{d\mu_i}{d\eta_i} x_{ir}x_{is}},\\
& \kappa_{\phi\phi}^{(\phi r)} =-\sum_{i=1}^n{\frac{\partial s_i}{\partial\mu_i}\frac{d\mu_i}{d\eta_i} x_{ir}},\\
& \kappa_{\phi\phi}^{(\phi \phi)} =-\sum_{i=1}^n{\frac{\partial s_i}{\partial\phi}}.\\
\end{align*}
%


\bibliographystyle{elsarticle-harv}
\bibliography{betabartlett}







\end{document}